\documentclass[aps,prd,reprint,onecolumn,superscriptaddress,nofootinbib]{revtex4-2}

\usepackage{graphicx}
\usepackage{amsmath}
\usepackage{amssymb}
\usepackage{hyperref}
\usepackage{cleveref}
\usepackage{float}
\usepackage{multirow}
\usepackage{siunitx}
\usepackage{bm}
\usepackage{xcolor}
\usepackage{booktabs}
\usepackage{array}
\usepackage{setspace}
\usepackage{geometry}
\usepackage{subcaption}
\RequirePackage{mathptmx}
\RequirePackage{flushend}
\usepackage{tabularx}
\usepackage{array}
\usepackage{booktabs}
\usepackage{lipsum}
\usepackage[utf8]{inputenc}
\usepackage{amsmath,amssymb}
\usepackage{graphicx}
\usepackage{array}
\usepackage{booktabs}
\usepackage{multirow}
\usepackage{xcolor}
\usepackage{hyperref}

\begin{document}

\title{Models for differential cross section in proton-proton scattering and their implications at ISR and LHC energies}
	
\author{Muhammad Saad Ashraf}
	
\author{Nosheen Akbar}
	
\affiliation{Department of Physics, COMSATS University Islamabad, Lahore Campus, Lahore, Pakistan.}
	
\author{Sarwat Zahra}
	
\affiliation{Department of Physics, DSNT, University of Education, Lahore, Pakistan.}
	
\email{msaadashraf99@gmail.com}
\email{nosheenakbar@cuilahore.edu.pk (corresponding author)}
\email{sarwat.zahra@ue.edu.pk}

\begin{abstract}

A few composite exponential models for the differential cross section ($\frac{d\sigma}{d|t|}$) are proposed to analyse the proton-proton ($pp$) elastic scattering at several energies. The parameters of these proposed models are found by fitting these models to the data for $pp$ elastic differential cross section reported at CERN-ISR, LHC, and extrapolated energies of other models. Plots of the data have important features, including dip-bump structure and shrinkage of the forward peak. Dips produced by our proposed models overlap the dips produced by data for a broad energy range of $\sqrt{s}$ = 23 GeV, 23.5 GeV, 27.23 GeV, 30.7 GeV, 44.7 GeV, 52.8 GeV, 62.5 GeV, 200 GeV, 800 GeV, 2.76 TeV, 7 TeV, 8 TeV, 13 TeV, 14 TeV, 15 TeV, and 28 TeV.
Employing these proposed models, elastic cross section ${(\sigma_{\text{el}})}$, inelastic cross section ${(\sigma_{\text{inel}})}$, and total cross section ${(\sigma_{\text{tot}})}$ are calculated at all the energies. Calculated results are compared with experimental data and theoretical results of other models. Implications of these results (obtained by models) related to the structure and dynamics of the proton are discussed. The findings of this study emphasize the significance of combining theoretical and phenomenological approaches to accurately describe $pp$ elastic scattering at high energies and provide significant information to future LHC experiments for the investigation of the differential cross section.

\singlespacing
\textbf{Keywords:} Elastic scattering, differential cross section, total elastic cross section, total cross section, inelastic cross section, QCD
		
\end{abstract}
	
\maketitle
\section{Introduction}
\label{sec:intro}
	When high-energy hadron-hadron collisions take place, either new particles are produced, or they scatter elastically ($h_1 h_2\rightarrow h_1 h_2$) with their quantum numbers conserved and without any production of new particles. In this study, we consider the latter case specified to proton-proton ($pp$) elastic scattering ($pp\rightarrow pp$). The $pp$ scattering process has its foundational role in high-energy particle physics. It provides a unique and controlled laboratory to probe the non-perturbative regime of Quantum Chromodynamics (QCD). This process is governed by strong interactions in principle and should be described by QCD. However, experimental data showed that the main features originate from non-perturbative QCD. In the absence of rigorous solutions of QCD equations, our approach to high-energy hadronic processes at present is still at an early stage of development. To gain an understanding of these hadronic processes, specifically the $pp$ elastic scattering, various phenomenological approaches have been developed based on Regge Theory. Precise and consistent parametrization of elastic scattering data enhances our understanding of hadron structure and the emergence of collective phenomena like gluon saturation, making a vital area of research in both theoretical and experimental high-energy physics. Study of $pp$ elastic scattering allows researchers to have direct access to the spatial and dynamical structure of the incoming protons through the measurement of the elastic differential cross section $\frac{d\sigma}{d|t|}$. The physical quantity $\frac{d\sigma}{d|t|}$ is among the most important quantities that have been calculated both experimentally and theoretically. It has been modeled in various contexts to describe the experimental data on $pp$ elastic differential cross section so that significant insights can be obtained, which can establish possible physical linkages with theoretical frameworks. The modeling of $\frac{d\sigma}{d|t|}$ is not only essential for interpreting collider data at CERN-Intersecting Storage Rings (ISR) and Large Hadron Collider (LHC) energies, but it also serves as a critical input for cosmic ray physics, hadronic Monte Carlo generators, and the development of theoretical frameworks such as eikonal models, Regge theory, and QCD-driven saturation approaches.
	
	\subsection{Experimental and Theoretical Contexts}
	Important ${pp}$ elastic scattering measurements are carried out by CERN-ISR at different center-of-mass energies in GeV in the 1970s. Among the most significant characteristics of differential cross section that they observed are the diffractive peak, the dip-bump structure, and the gradual increase in the total cross section with increase in ${\sqrt{s}}$. Soft diffractive processes resulting from pomeron, reggeon, and meson exchanges are responsible for elastic scattering at low energies. According to ref. \cite{DONNACHIE1992227}, soft diffractive interactions controlled by pomeron exchange within the Regge regime dominate elastic scattering at low momentum ${\mid t\mid\leq 0.1}$ ${\textrm{GeV}^{2}}$. A change from soft pomeron-dominated scattering to the perturbative QCD (pQCD) regime at intermediate momentum ${\mid t\mid\sim1-2}$ ${\textrm{GeV}^{2}}$ \cite{PhysRevD.83.077901} is indicated by deviations from simple exponential fits. At high momentum ${\mid t\mid > 2}$ ${\textrm{GeV}^{2}}$, perturbative QCD effects and possible indications of parton saturation or color transparency become significant beyond the dip region, which can be understood from refs. \cite{DONNACHIE2013500,Donnachie:2002en,Forshaw:1997dc}. For a broad momentum range of ${\mid t\mid\approx}$ 0.8-10 ${\textrm{GeV}^{2}}$, the differential cross sections for $\sqrt{s}$ = 23-63 ${\textrm{GeV}}$ at ISR energies have been measured in ref. \cite{NAGY1979221}.
	
	Among the important advances achieved by the TOTEM Collaboration at LHC energies, the accurate measurements of ${pp}$ elastic scattering are at ${\sqrt{s}}$ = 2.76, 7, 8, 13 TeV under different experimental settings as described in refs. \cite{TOTEM_2018psk,Antchev_2013,Antchev_2011,TOTEM:2021imi,Antchev_2019}. The sequence of TOTEM measurements revealed that the diffractive minimum and diffractive maximum, also known as dip and bump, are a continuous feature of the ${pp}$ differential cross-section, in contrast to ${\bar{p}p}$ at high energies. By comparing ${pp}$ and ${\bar{p}p}$ interactions, the D0 and TOTEM Collaborations recently discovered the C-odd exchange, sometimes referred to as the Odderon exchange \cite{PhysRevLett.127.062003}. High-precision measurements of the $\rho$ parameter, elastic and total cross-sections have been made by the TOTEM Collaboration, revealing important aspects of elastic scattering at TeV energies. Other significant TeV-scale measurements done prior to TOTEM include the UA4 collaboration \cite{Bozzo1984LowMT} \cite{UA41985}, CDF collaboration \cite{CDF1994}, E710 collaboration \cite{E7101988, E7101989, E7101992}, ATLAS experiment \cite{ATLAS2016}, and RHIC \cite{STAR2016}. The vast amount of data gathered from these measurements has led to the development of several theoretical and phenomenological models \cite{Froissart1961, Gauron1992, Block1994, Cudell2002, COMPETE2002}. However, the new TOTEM data from ${\sqrt{s}}$ = 2.76 to 13 TeV \cite{TOTEM2018} is not sufficiently explained by these models.
	
	One of the most effective methods for examining the dynamical and spatial structure of the proton is the study of elastic $pp$ scattering. Elastic processes give direct access to the spatial distribution of matter within the proton and contribute around one-fourth of the total cross section at high energy. According to ref.\cite{MMIslam}, the proton has a layered macrostructure: 1) a dense inner core made up of confined valence quarks and gluons; 2) a middle shell related to the distribution of baryonic charge; and 3) an outer peripheral region dominated by meson cloud contributions (\(q\bar{q}\) condensates). In elastic collisions, the momentum transfer is related to the transverse distance $b$ by $b \sim 1/\sqrt{|t|}$. Thus, higher {$\mid t\mid$} values probe deeper into the baryonic core and the valence quark region, whereas low-${\mid t\mid}$ elastic scattering probes the large-distance, low-density mesonic cloud. When momentum transfers are low, the differential cross section $\frac{d\sigma}{dt} \simeq e^{Bt}$ with $B$ as the slope parameter describing the transverse extension of the interaction region, exhibits approximately exponential behavior. The deviations from pure exponential behavior become evident as \(t\) grows, suggesting the inclusion of interference effects between various spatial components and internal proton layers. Quark-quark scattering dominates the deep elastic scattering domain above \(|t| \gtrsim 4\,\mathrm{GeV}^2\).
	
	As the center of mass energy is raised from ISR to LHC energies, the proton becomes more absorbent and blacker, its interaction profile becomes more peripheral, and its effective interaction radius grows \cite{DREMIN2013241}. At ISR energies, a relatively compact proton interaction region is dominated by valence quarks and sea quarks with a moderate density of gluons. As energy increases approaching LHC scales, the interaction radius ($R_{\text{int}}$) increases due to the growth of gluon densities at small Bjorken-$x$. As the slope parameter $B(s)$ rises from around 14 \textrm{GeV}$^{-2}$ (ISR) to over 21.19 \textrm{GeV}$^{-2}$ (LHC), this growth is evident in the exponential fit to the elastic cross-section \cite{PhysRevD.108.034028}. The "shrinkage of the forward peak" and the deepening and shifting of the diffractive dip in \(d\sigma/dt\) towards lower \(|t|\) values are the empirically observed manifestations of it. The gluon density within the proton increases considerably at LHC energy, resulting in nonlinear QCD processes such as gluon recombination and saturation \cite{Gelis2010}. At ISR energies, there is uniformity in the impact-parameter profile of the proton. On the other hand, different core-corona structures are formed at LHC energies due to double-parton interactions. The proton is frequently described as a spherically symmetric particle at ISR, but at LHC, it takes on an asymmetric form due to event-by-event fluctuations in the gluon distributions that depend on the impact parameter.
	
	In recent decades, researchers have proposed a wide range of theoretical and phenomenological models to represent ${pp}$ elastic scattering data ranging from ISR and LHC energies. However, more sophisticated models have been required to achieve full agreement with experimental results. The energy dependence of total cross-sections and elastic scattering employing pomeron exchange is described in ref.\cite{DONNACHIE1992227}. The non-exponential behavior in the differential cross-section at low ${|t|}$, however, is not amenable to fitting by traditional Regge models. In ref.\cite{Saleem:1980hu}, the dip-bump structure in the differential cross section is described only at low energies using a dipole pomeron-based formalism. In $pp$ and $p\bar{p}$ elastic scattering at low momentum transfers, the interplay between electromagnetic and hadronic interactions has been extensively studied using the Coulomb-Nuclear Interference (CNI) model, which was developed by Cahn in 1980 \cite{Cahn:1980}. The CNI model, which included nuclear and Coulomb amplitudes in an eikonal formalism, offered a foundation for calculating the hadronic phase shift. The CNI model is further utilized and improved in ref. \cite{Kundrat:1994}. At ISR and LHC energies, limitations of the CNI model related to the Coulomb phase, spin effects, and nonlinear QCD corrections also exist. In order to better fit the experimental results, the TOTEM Collaboration in \cite{TOT2015527} employed quadratic and cubic polynomial parameterizations in the exponent at ${\sqrt{s}}$= 8 TeV in the range of ${0.027 < {\mid t \mid} < 0.2}$ \textrm{GeV}². A number of studies that employed these parametrizations are included in the refs. \cite{TOTEM:2019,Kohara:2017aix,Jenkovszky:2018pcm}. In ref. \cite{MARTYNOV2018414}, several parametrizations are developed to explain the differences in between ${pp}$ and ${\bar{p}p}$ scattering and provided qualitative agreement for exchange of Odderon. At high energies, the gluon density inside the proton grows significantly, and saturation models quite explicitly describe this regime. A QCD-based framework that generalizes parton distributions (GPDs) extends to both soft and hard interactions, which is provided by the HEGS (High Energy Generalized Structure) for ${pp}$ elastic scattering refs.\cite{HEGS1, HEGS2}. According to refs.\cite{HEGS4, HEGS5}, the model described two-gluon and three-gluon exchange contributions and matched the ISR and LHC data, as well as the ${\sigma_{tot}(s)}$ and ${B(s)}$. The HEGS model still has to be refined in terms of unitarization and Odderon contributions to accurately describe the scattering amplitude. In ref. \cite{HoloQCD1}, elastic ${pp}$ scattering in the Regge regime is investigated in a holographic QCD model, and consistent outcomes have been obtained using 13 TeV TOTEM data. An empirical model that is developed and proved to be in good agreement with LHC and ISR data is based on the early work of Phillips and Barger \cite{PHILLIPS1973412}. An exponential shape of the differential cross-section is assumed by the Phillips-Barger Model (PB), which may be used to determine the parton distribution inside the proton. The recent ${pp}$ and ${\bar{p}p}$ elastic scattering data at ${\sqrt{s}}$ = 8 and 13 TeV are represented by an improved Phillips-Barger model in ref.\cite{Gonçalves1973}, which also highlighted the significance of the Odderon exchange. Elastic scattering is described at ISR energies \cite{Nemes:2015iia} and at an LHC energy of ${\sqrt{s}}$ = 7 TeV \cite{Csorgo2023} by modifying the Bialas-Bzdak model, which considered the proton as a bound state of a quark and a diquark. A constant bump-to-dip ratio of the differential cross-section $\frac{d\sigma_{el}}{dt}$ at the ISR and LHC energies is found in the context of geometrical scaling in ref. \cite{Praszalowicz:2025djk}. The constant bump-to-dip ratio and a family of scaling rules are also found at the LHC in ref. \cite{BALDENEGRO2024138960}.
	
	A precise parametrization and fitting of the differential cross-section in ${pp}$ elastic scattering at ISR and LHC energies has important implications for hadronic interactions, QCD dynamics, and proton structure. The Coulomb-Nuclear Interference (CNI) phase, which is essential for understanding low-${\mid t\mid}$ behavior, the elastic slope parameter B(s), which probes the interaction range, and the total cross-section through the Optical Theorem can all be extracted using these expressions \cite{DONNACHIE1992227}\cite{TOTEM2018}\cite{Fagundes22013}\cite{Block2011}. Fits can reveal information on the spatial distribution of the interaction region of the proton, demonstrating an energy-dependent increase in the radius of the proton, through the slope parameter ${B(s)}$. By using Fourier transforms, these parametrizations allow for developing a model of the impact parameter distribution, which can provide information about the spatial structure of proton \cite{Anisovich2014}. Additionally, refined Parton Distribution Functions (PDFs) \cite{Dulat2016} improve collider luminosity calibration \cite{TOTEM2018, Antchev2013}, and are vital for high-energy cosmic-ray interactions \cite{Auger2015, Ostapchenko2019}. They make it easier to examine non-perturbative QCD models, such as saturation and eikonal frameworks \cite{Grau2018}\cite{Dremin20131}, and to extrapolate cross-sections to future collider energies, such as the FCC \cite{FCC2021}. Applications such as astroparticle research, neutrino physics, and machine learning-based cross-section predictions can all benefit from these parameterizations \cite{Basso2021}. For the advancement of theoretical and experimental high-energy physics and the investigation of strong interaction dynamics, continuous advancement is crucial. It is emphasized that the only way to provide substantial support for theoretical advancements is to conduct a global comparative analysis of various methodologies, parametrizations, and solutions to obtain empirical data on what is universal across every context. The offered method and results aim to advance research into inverse problems in high-energy elastic hadron-hadron scattering.
	
	To explain the recent ${pp}$ elastic scattering measurements by the TOTEM Collaboration at LHC and previously by the CERN Intersecting Storage Rings (ISR) and their extrapolations, we present several models of elastic differential cross section $\frac{d\sigma}{d|t|}$ as a function of $s$ and ${|t|}$, the magnitude of the four-momentum transfer squared based on empirical observations.
	
	We describe our methodical approach towards modeling the elastic differential cross section $\frac{d\sigma}{d|t|}$ in Section \uppercase{ii}. The results of fitting our models to all the data are shown in Section \uppercase{iii}. The calculations of the elastic, total, and inelastic cross sections by our models are written in this section. The $\chi^{2}$ and other error measures for differential cross section model fitting and total elastic cross section are also written in Section \uppercase{iii}. In Section \uppercase{iii}, the potential physical significance of our models and how they could relate to other models is also explored. Section \uppercase{iv} provides the conclusion of this study. Additionally, Section \uppercase{v} offers recommendations for future investigations in the directions highlighted by this study.

	\section{Methodology}

	In this study, seven different models based on their ability to reproduce key features of the differential cross-section, including the forward peak, diffractive dip, and large-${\mid t\mid}$ behavior, are developed to fit the ISR and TOTEM elastic ${pp}$ differential cross-section data. The ISR data at low energies: ${\sqrt{s}}$ = 23, 23.5, 30.7, 44.7, 52.8, 62.5 \text{GeV} covered a broad range of momentum transfer squared ${0 < {\mid t\mid} < 9.75}$ ${\textrm{GeV}^{2}}$ refs. \cite{NAGY1979221}\cite{Amaldi1980}. The data from TOTEM experiment at higher energies: ${\sqrt{s}}$ = 2.76, 7, 8, 13 \text{ TeV} from refs. \cite{TOTEM_2018psk,Antchev_2013,Antchev_2011,TOTEM:2021imi,Antchev_2019}, contained precise results of the differential cross-section at a low range of ${0.037 < {\mid t\mid} < 4.03}$ ${\textrm{GeV}^{2}}$ values. Non-linear fitting for the models is performed for both datasets, and the parameters of the models of the elastic differential cross section are determined. The models are also fitted to extrapolated results at future LHC energies of 14, 15, and 28 TeV of ref. \cite{Nemes:2015iia}. Predicted results at 200 GeV and 800 GeV of ref. \cite{Bourrely:1978da}, and at 27.43 GeV of \cite{Bence:2020usl} are also fitted by these models. For the non-linear fitting, the available experimental and extrapolated data of the differential cross section are taken only in the ${\mid t\mid}$ ranges given in Table I. In these ranges, the data essentially contains the dip-bump structure and most part of the large-${\mid t\mid}$ tail.
	
	\renewcommand{\arraystretch}{1.5}
	\setlength{\tabcolsep}{8pt}
	\begin{table}[h]
		\centering
		\caption{Ranges of $\mid t \mid$ for the data at each $\sqrt{s}$ value.}
		\label{tab:parameters}
		\scalebox{0.80}{
			\begin{tabular}{c|c|c|c}
				\toprule
				$\sqrt{s}$ & $|t|$ & $\sqrt{s}$ & $|t|$ \\
				\midrule
				${\textrm{GeV}}$ & ${\textrm{GeV}^{2}}$ & ${\textrm{TeV}}$ & ${\textrm{GeV}^{2}}$ \\
				\midrule
				23  & $0.04 \leq {\mid t\mid} \leq 3.6$ & 2.76 & $0.3825 \leq {\mid t\mid} \leq 0.7625$ \\
				23.5 & $0.042 \leq {\mid t\mid} \leq 5.75$  & 7 & $0.381 \leq {\mid t\mid} \leq 2.543$ \\
				27.43  & $0.4540 \leq {\mid t\mid} \leq 9.8772$ & 8  & $0.2078 \leq {\mid t\mid} \leq 1.8647$ \\
				30.7  & $0.0011 \leq {\mid t\mid} \leq 5.75$ & 13 & $0.0393 \leq {\mid t\mid} \leq 4.0346$ \\
				44.7  & $0.0010 \leq {\mid t\mid} \leq 7.25$ & 14 & $0.0071 \leq {\mid t\mid} \leq 2.1413$ \\
				52.8  & $0.0013 \leq {\mid t\mid} \leq 9.75$ & 15 & $0.0072 \leq {\mid t\mid} \leq 2.4713$ \\
				62.5  & $0.0017 \leq {\mid t\mid} \leq 6.25$ & 28 & $0.0143 \leq {\mid t\mid} \leq 2.4786$ \\
				200 & $0.0263 \leq {\mid t\mid} \leq 9.8109$ &  &  \\
				800 & $0.0060 \leq {\mid t\mid} \leq 9.9706$ &  & \\
				\bottomrule
			\end{tabular}
		}
	\end{table}
 	
\subsection{Models for the Differential Cross Section}
Different parametrizations, which are constructed as multi-component exponential structures for modeling the differential cross section for ${pp}$ elastic scattering, are presented here. These models represent the composite behavior of the scattering amplitude in different kinematic regimes to fit the data. In these models, scaling-like factors $(s/s_0)$ have been introduced with dimensionless exponents, $\eta$ and $\mu$, to fit the data. In this factor, the quantity $s_0$ is taken to be of the order of a $\textrm{GeV}^{2}$, because this is the typical scale associated with high-energy hadronic reactions. Such s dependent factors have been used to approximate the differential cross sections of various scattering phenomena in ref. \cite{Brodsky:1973kr}. These factors are also reminiscent of the low-$\mid t\mid$ ($\mid t\mid < 1 \textrm{GeV}^{2}$) form of the $pp$ and $p\bar{p}$ elastic differential cross section model of ref. \cite{Donnachie:2002en} in which single pomeron ($\mathbb{P}$) and double pomeron exchanges ($\mathbb{P}\mathbb{P}$) have been expressed in terms of their trajectories. This model describes $pp$ elastic scattering in terms of pomeron coupling with the separate valence quarks of the proton. For better understanding and correlation purposes with this low-$\mid t\mid$ model, we mention its expression as follows
\begin{equation}
	\frac{d\sigma}{dt} = (\frac{(3 \beta_P F_1(t))^{4}}{4\pi})(s/s_0)^{{2\alpha_P }(t)-2}.
\end{equation}
The proton form factor $F_1 (t)$ is parameterized as $(F_1 (t))^{2} = A e^{a t} + B e^{b t} + C e^{c t} $ for calculational convenience and the trajectory for the pomeron exchange is given as $\alpha_P (t) = 1 + \epsilon_P + \alpha_P ' t. $ The constant $\beta_P$ originated from the pomeron-exchange contribution to quark-quark scattering. Keeping these facts in view, we have introduced simple energy-dependent generalized parameters (A(s) and B(s)) that will appear in our models as $A(s) = A_0 (s/s_0)^{\eta}$ and $B(s) = B_0 (s/s_0)^{\mu}$ respectively. This form not only offers a smaller number of parameters but also reduces statistical errors in fitting. Each model is distinct and involves several free parameters that are determined by fitting to all the data. In this study, the analysis of the data is performed by non-linear fitting; the final outcome profoundly depends on the choice of initial values of the free parameters, which are not known. Therefore, reaching a unique solution is not certain, but only one or more possible solutions.The most optimal set of parameter values for each model across a center-of-mass energy is the set of values for which the curve fitting and the calculations can show the most agreement with experimental measurements and extrapolated prediction where available. Our models differ from traditional Regge models \cite{Jenkovzsky,Pancheri1}, which usually involve many parameters. We now describe our models as follows.

\subsubsection{Model 1}

Model 1 is introduced with four exponential terms in the following form:
\begin{equation}
	\frac{d\sigma (s,t)}{dt} = (s/s_0)^\eta (A_1 e^{-B_1 (s/s_0)^\mu t} + A_2 e^{-B_2 (s/s_0)^\mu t} + A_3 e^{-B_3 (s/s_0)^\mu t} + A_4 e^{-B_4 (s/s_0)^\mu t}).
\end{equation}
Here, the parameters  ${A_1}$, ${A_2}$, ${A_3}$, ${A_4}$, ${B_1}$, ${B_2}$, ${B_3}$, ${B_4}$, $\eta$ and $\mu$ are treated as free parameters which are determined through fitting this model with $pp$ elastic differential cross section data for the energies mentioned in Table I. The values of all the parameters of this model are given in Table II.

\subsubsection{Model 2}
This model is considered in the following form.
\begin{equation}
	\frac{d\sigma (s,t)}{dt} = (s/s_0)^\eta (A_1 e^{-B_1 (s/s_0)^\mu t}(1+\alpha e^{-C_1 (s/s_0)^\mu t}) + A_2 e^{-B_2 (s/s_0)^\mu t} + A_3e^{-B_3 (s/s_0)^\mu t} + A_4 e^{-B_4 (s/s_0)^\mu t} t^{-m}).
\end{equation}
Here the parameters ${A_1}$, ${A_2}$, ${A_3}$, ${A_4}$, ${B_1}$, ${B_2}$, ${B_3}$, ${B_4}$, ${C_1}$, $\alpha$, $\eta$, $m$, and $\mu$ are treated as free parameters which are determined through fitting this model with $pp$ elastic differential cross section data for the energies mentioned in Table I. The first two terms have the same amplitude coefficient ${A_1}$ and the $e^{-B_1 (s/s_0)^\mu t}$ factors. A power law factor $t^{-m}$ is combined with exponential damping in the last term for better fitting. This factor allows the model to smoothly transition from soft diffraction (low-${\mid t\mid}$) to hard scattering behavior (high-${\mid t\mid}$). The values of all the parameters of this model are given in Table III.

\subsubsection{Model 3}
This exponential model with a quadratic modulation factor $(1-\alpha t/t_0)^{2}$ in the second exponential term is written as:
\begin{equation}
	\frac{d\sigma (s,t)}{dt} = (s/s_0)^\eta (A_1 e^{-B_1 (s/s_0)^\mu t} + A_2 e^{-B_2 (s/s_0)^\mu t}(1-\alpha t/t_0)^{2} + A_3 e^{-B_3 (s/s_0)^\mu t} + A_4e^{-B_4 (s/s_0)^\mu t} t^{-m}).
\end{equation}
The quadratic modulation factor is zero for $t=t_0/\alpha$ and equal to 1 for t=0. Therefore, the exponential part of this term dominates at $t=0$. As t increases, the polynomial suppresses the exponential behavior depending on the sign of ${\alpha}$, thus affecting the dip-bump region. At high t values, this factor is useful to significantly suppress or enhance the tail behavior by changing the low-${\mid t\mid}$ fit. The last exponential term contains a power law factor ${t^{-m}}$. For small t, ${t^{-m}}$ can be large if $m>0$, but the exponential term may dominate when $B_4 (s/s_0)^\mu t \rightarrow 0$. For large t, ${t^{-m}}$ becomes small, but this may slow the overall decay if the exponential is decreasing too fast. The free parameters ${A_1}$, ${A_2}$, ${A_3}$, ${A_4}$, ${B_1}$, ${B_2}$, ${B_3}$, ${B_4}$, $\alpha$, $m$, $\eta$, and $\mu$ are determined through fitting this model with $pp$ elastic differential cross section data for the energies mentioned in Table I. The values of all the parameters of this model are given in Table IV.

\subsubsection{Model 4}
This exponential model with two quadratic modulation factors is written in the following form:
\begin{equation}
	\frac{d\sigma (s,t)}{dt} = (s/s_0)^\eta (A_1 e^{-B_1 (s/s_0)^\mu t}(1-\alpha t/t_0)^{2} + A_2 e^{-B_2 (s/s_0)^\mu t} (1-\beta t/t_0)^{2} + A_3 e^{-B_3 (s/s_0)^\mu t} + A_4e^{-B_4 (s/s_0)^\mu t}).
\end{equation}
This model contains mainly four exponential terms with two quadratic modulation factors, $(1-\alpha t/t_0)^{2}$ and $(1-\beta t/t_0)^{2}$ multiplied with the first and second terms, respectively. These factors become zero when $t=t_0/\alpha$ and $t=t_0/\beta$. When $t=0$, these factors are one and preserve the exponential factors. The values of all the free parameters (${A_1}$, ${A_2}$, ${A_3}$, ${A_4}$, ${B_1}$, ${B_2}$, ${B_3}$, ${B_4}$, $\alpha$, $\beta$, $t_0$, $\eta$, $\mu$) of this model are found by fitting this model with $pp$ elastic differential cross section data for the energies mentioned in Table I. Values of parameters are given in Table V.

\subsubsection{Model 5}
This exponential model with saturation-like correction factor is given in the following form
\begin{equation}
	\frac{d\sigma (s,t)}{dt} = (s/s_0)^\eta (A_1 e^{-B_1 (s/s_0)^\mu t}t^{-m} + A_2 e^{-B_2 (s/s_0)^\mu t}t^{-n} + A_3 e^{-B_3 (s/s_0)^\mu t}(1-t/t_0)^p + A_4e^{-B_4 (s/s_0)^\mu t}).
\end{equation}
The values of all the parameters are found by fitting this model the model with $pp$ elastic differential cross section data for the energies mentioned in Table I, and the values of these parameters are given in Table VI. The parameter $t_0$ is a reference momentum scale (in $\textrm{GeV}^{2}$) controlling the onset of the saturation term. The first two terms combine exponential damping with power-law behavior, fitting the forward peak and early falloff at low and intermediate t values, respectively. The third term contains a saturation-like correction factor $(1-t/t_0)^p$ which becomes significant at the mid-t region and helps to describe the dip-bump structure at higher energies. The last exponential term accounts for the large-t tail of the cross section as it provides a smooth decay at high t.

\subsubsection{Model 6}
This exponential model with logarithmic correction factor is given in the following form:
\begin{equation}
	\frac{d\sigma (s,t)}{dt} = (s/s_0)^\eta (A_1 e^{-B_1 (s/s_0)^\mu t}t^{-m} + A_2 e^{-B_2 (s/s_0)^\mu t}t^{-n}Log(1+\alpha t/t_0) + A_3 e^{-B_3 (s/s_0)^\mu t} + A_4e^{-B_4 (s/s_0)^\mu t}).
\end{equation}
The parameters ${A_1}$, ${A_2}$, ${A_3}$, ${A_4}$, ${B_1}$, ${B_2}$, ${B_3}$, ${B_4}$, $\alpha$, $\eta$, $\mu$, $m$, and $n$ are free parameters, determined through fitting the model with $pp$ elastic differential cross section data for the energies mentioned in Table I. The parameter $t_0$ is a reference momentum scale (in $\textrm{GeV}^{2}$) controlling the onset of the saturation term. However, the first term contains a power law factor ${t^{-m}}$ which suppresses the cross section at high t values. And the second exponential term contains a power law factor ${t^{-n}}$ and a logarithmic correction $Log(1+\alpha t/t_0)$. The logarithmic factor grows slowly and enhances the behavior of the cross section mildly as t increases. At small t, $Log(1+\alpha t/t_0) \approx \alpha t/t_0$ and it behaves linearly, i.e., the whole correction $\approx t^{-(n-1)}$. At large t, the logarithm grows slowly, thereby adding a soft enhancement to the suppression. The parameter values of this model are given in Table VII.

\subsubsection{Model 7}
This exponential model with stretched exponential and power-law-type factors is given in the following form:
\begin{equation}
	\frac{d\sigma (s,t)}{dt} = (s/s_0)^\eta (A_1 e^{-B_1 (s/s_0)^\mu t} + A_2 e^{-B_2 (s/s_0)^\mu t} + A_3 e^{-B_3 (s/s_0)^\mu t} + A_4e^{-B_4 (s/s_0)^\mu t})e^{-\gamma (t/t_0)^{m}}(1+t/t_0)^{-n}.
\end{equation}
Here ${A_1}$, ${A_2}$, ${A_3}$, ${A_4}$, ${B_1}$, ${B_2}$, ${B_3}$, ${B_4}$, $\eta$, $\mu$, $\gamma$, $m$, and $n$ are free parameters, determined through fitting this model with $pp$ elastic differential cross section data for the energies mentioned in Table I. The parameter $t_0$ is a reference momentum scale (in $\textrm{GeV}^{2}$). In this model, each term is multiplied by two factors: $e^{-\gamma (t/t_0)^{m}}$, a stretched exponential, and $(1+t/t_0)^{-n}$, a power-law-type decay factor. Transition between the exponential regime at low t and the power-law-like regime is made smoother by these factors. Multiplying  $e^{-\gamma (t/t_0)^{m}}$ with every term modifies the slope beyond a certain range. If $m=2$, then it can act like a Gaussian suppression. If $m<1$, then it may decay more slowly than exponential. If $\gamma$ is large, then stronger suppression occurs in mid-to-high t. The power-law-like factor $(1+t/t_0)^{-n}$ gives additional control over the tail behavior at high t. For $t\gg t_0$, $(1+t/t_0)^{-n} \approx (t/t_0)^{-n}$ which provides high momentum suppression. The values of all the parameters of this model are given in Table VIII.

These proposed models are used to calculate the total elastic and inelastic cross sections as discussed in detail in Section \uppercase{iii}. Chi-square is calculated to determine the accuracy of the models. Absolute relative error (ARE) is calculated for the total elastic cross section. The calculated values of ${\sigma_{\text{el}}}$, ${\sigma_{\text{tot}}}$, and ${\sigma_{\text{inel}}}$ are compared with others' work, and the best models that predicted the most accurate values are identified.

	\section{Results and Discussion}
	\label{sec:results}
	In this study, we performed a systematic fitting of elastic differential cross-section data for $pp$ scattering over a wide range of energies, including both GeV and TeV scales, using seven distinct composite exponential models. Each model is designed to capture different kinematic features of the data, reflecting the complex internal structure of the proton and the energy-dependent nature of elastic scattering. Here, we discuss the results with emphasis on the possible physical insights extracted from the fitting behavior of our models with the data. The models are fitted to the elastic scattering data collected at various center-of-mass energies, from GeV energies of 23, 23.5, 27.43, 30.7, 44.7, 52.8, and 62.5 GeV to LHC energies of 2.76, 7, 8, and 13 TeV, and at extrapolated values at 200 GeV, 800 GeV, 14 TeV, 15 TeV, and 28 TeV. The results of the obtained fits are shown in Figures 1-7.
		\begin{figure}[h!]
		\centering
		\begin{subfigure}[b]{0.49\textwidth}
			\centering
			\includegraphics[width=\textwidth]{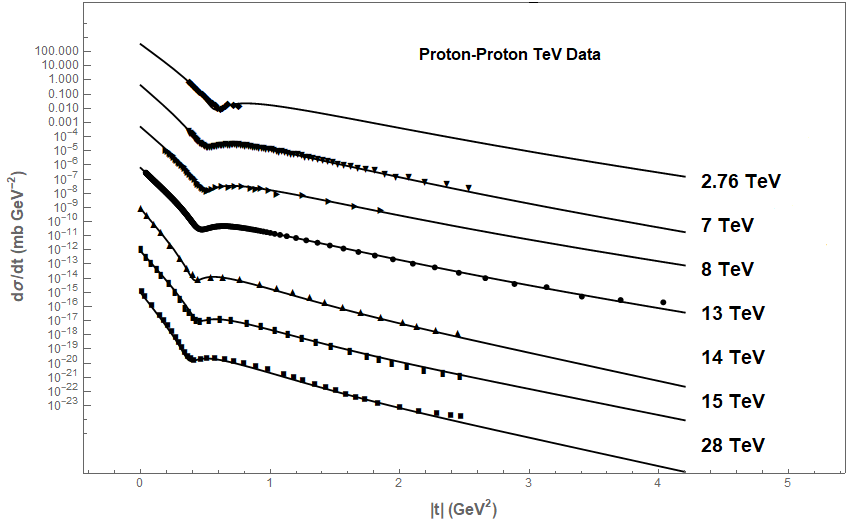}
			\caption{}
			\label{fig:a}
		\end{subfigure}
		\hfill
		\begin{subfigure}[b]{0.49\textwidth}
			\centering
			\includegraphics[width=\textwidth]{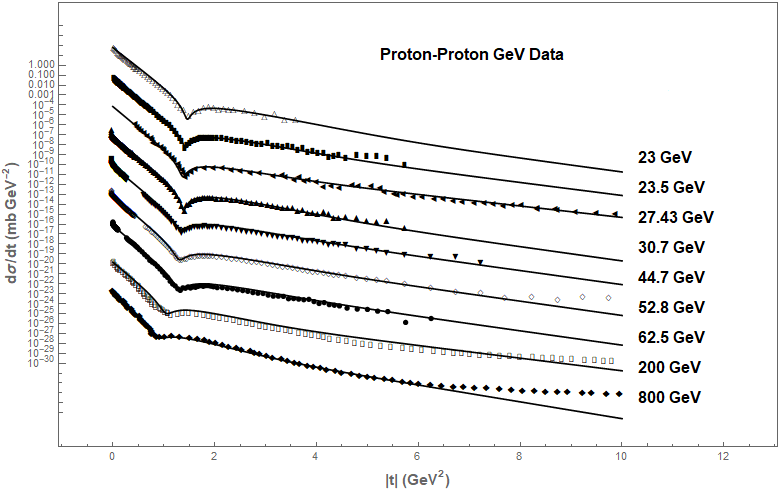}
			\caption{}
			\label{fig:b}
		\end{subfigure}
		\caption{(a) Model 1 fits on the $pp$ elastic differential cross section data in the energy range $2.76 \leq \sqrt{s} \leq 28  \textrm{TeV}$. Filled squares represent data points taken from extrapolated data at 28 TeV, filled rectangles represent data points taken from extrapolated data at 15 TeV, filled up triangles represent data points taken from extrapolated data at 14 TeV, filled circles represents the data at 13 TeV, filled right triangles represent the data at 8 TeV, filled down triangles represent the data 7 TeV, and filled diamonds represent the data at 2.76 TeV. Data and model values are multiplied by $10^{-3(n-1)}$, where $n$ is the number of curve and corresponding data set starting from the top. The solid line represents the fit of our model to the data.
			(b) Model 1 fits the $pp$ elastic scattering data of differential cross section in the energy range $23 \leq \sqrt{s} \leq 800  \textrm{GeV}$. Filled diamonds represent data points taken from extrapolated data at 800 GeV, empty rectangles represent the data at 200 GeV, filled circles represent data at 62.5 GeV, empty diamonds represent the data at 52.8 GeV, filled down triangles represent the data at 44.7 GeV, filled up triangles represent the data at 30.7 GeV, filled left triangles represent the data at 27.43 GeV, filled solid rectangles represent the data at 23.5 GeV, and empty up triangles represent the data at 23 GeV. Data and model values are multiplied by $10^{-3(n-1)}$, where $n$ is the number of curve and corresponding dataset starting from top. The solid line represents the fit of our model to the data.
		}
		\label{fig:combined}
	\end{figure}
	
		\begin{figure}[h!]
		\centering
		\begin{subfigure}[b]{0.49\textwidth}
			\centering
			\includegraphics[width=\textwidth]{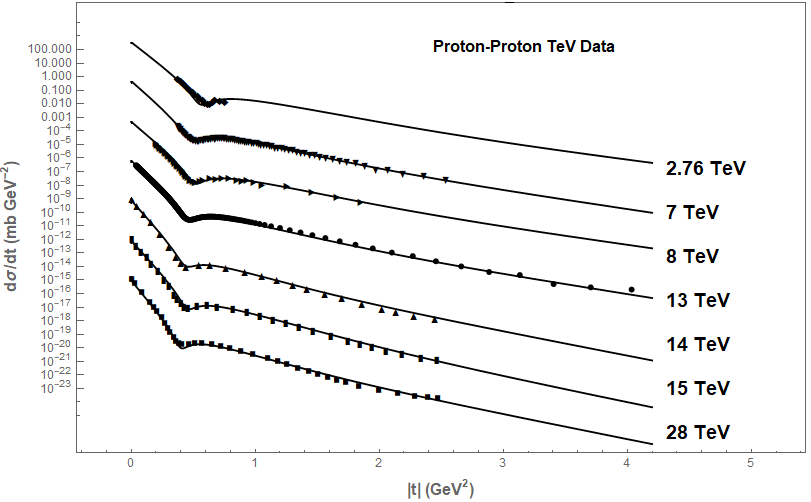}
			\caption{}
			\label{fig:a}
		\end{subfigure}
		\hfill
		\begin{subfigure}[b]{0.49\textwidth}
			\centering
			\includegraphics[width=\textwidth]{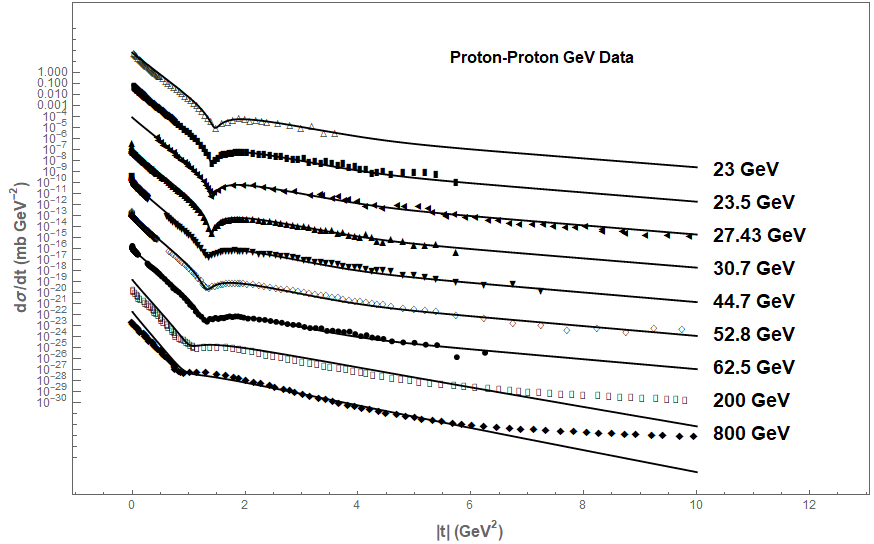}
			\caption{}
			\label{fig:b}
		\end{subfigure}
		\caption{(a) Same legend as in Fig.1 with Model 2. (b) Same legend as in Fig.1 with Model 2.}
		\label{fig:combined}
	\end{figure}
	\begin{figure}[h!]
		\centering
		\begin{subfigure}[b]{0.49\textwidth}
			\centering
			\includegraphics[width=\textwidth]{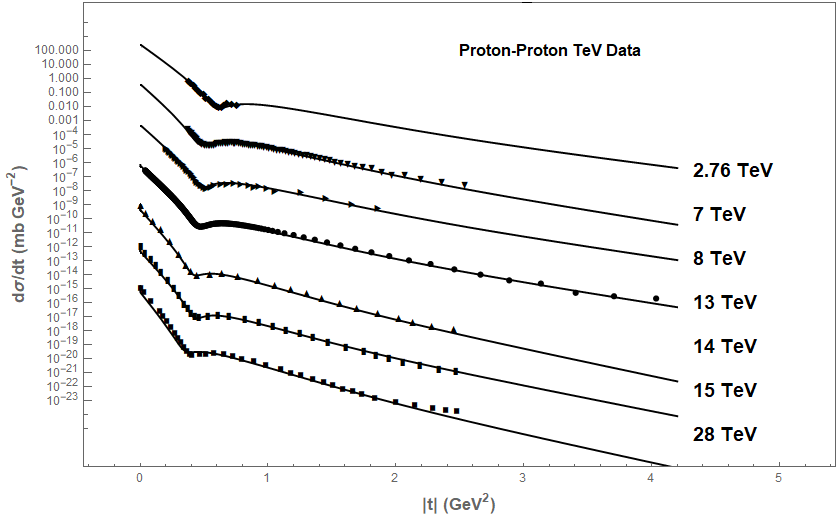}
			\caption{}
			\label{fig:a}
		\end{subfigure}
		\hfill
		\begin{subfigure}[b]{0.49\textwidth}
			\centering
			\includegraphics[width=\textwidth]{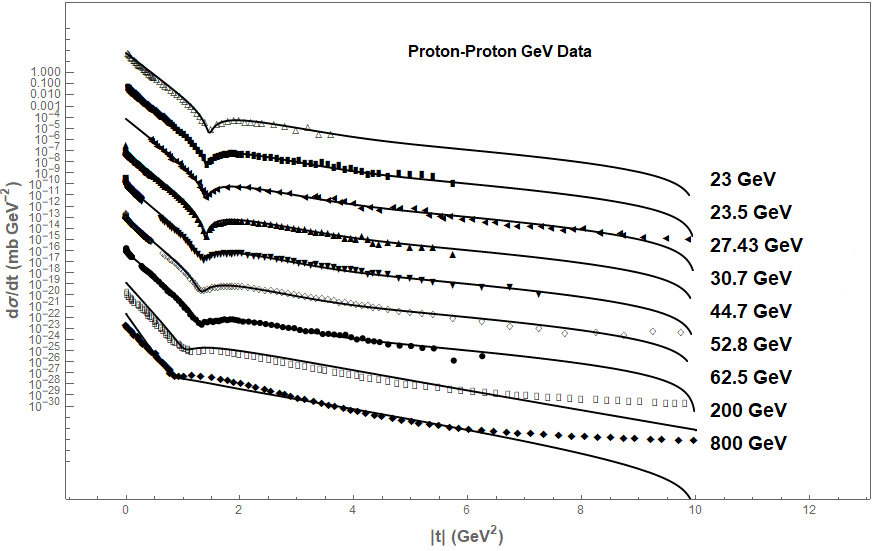}
			\caption{}
			\label{fig:b}
		\end{subfigure}
		\caption{(a) Same legend as in Fig.1 with Model 3. (b) Same legend as in Fig.1 with Model 3.}
		\label{fig:combined}
	\end{figure}
		\begin{figure}[h!]
		\centering
		\begin{subfigure}[b]{0.49\textwidth}
			\centering
			\includegraphics[width=\textwidth]{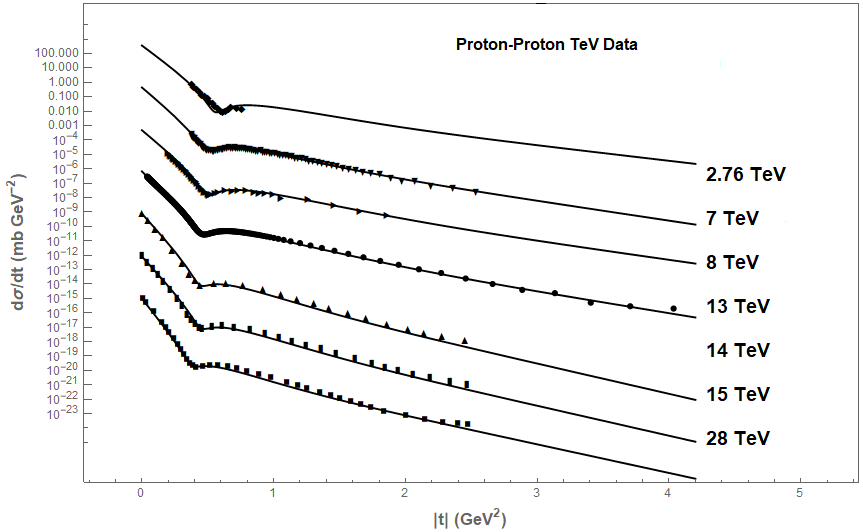}
			\caption{}
			\label{fig:a}
		\end{subfigure}
		\hfill
		\begin{subfigure}[b]{0.49\textwidth}
			\centering
			\includegraphics[width=\textwidth]{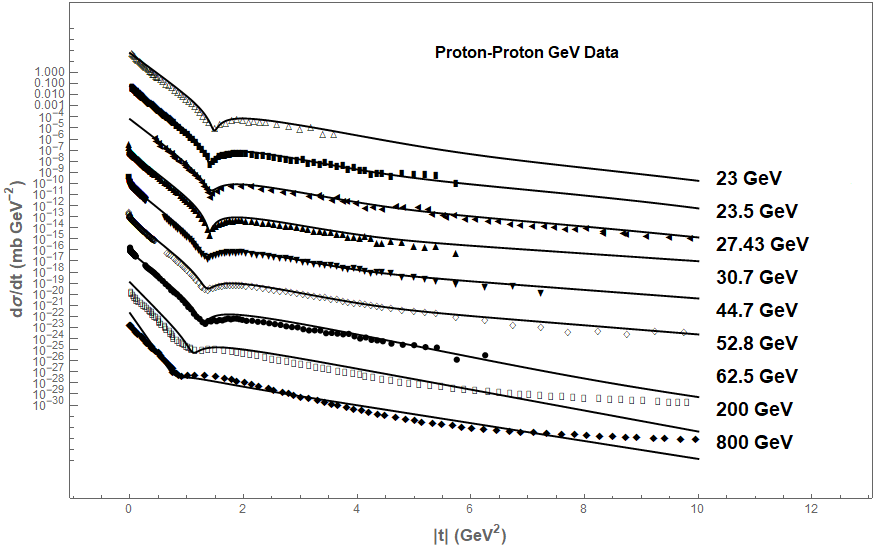}
			\caption{}
			\label{fig:b}
		\end{subfigure}
		\caption{(a) Same legend as in Fig.1 with Model 4. (b) Same legend as in Fig.1 with Model 4.}
		\label{fig:combined}
	\end{figure}
	\begin{figure}[h!]
		\centering
		\begin{subfigure}[b]{0.49\textwidth}
			\centering
			\includegraphics[width=\textwidth]{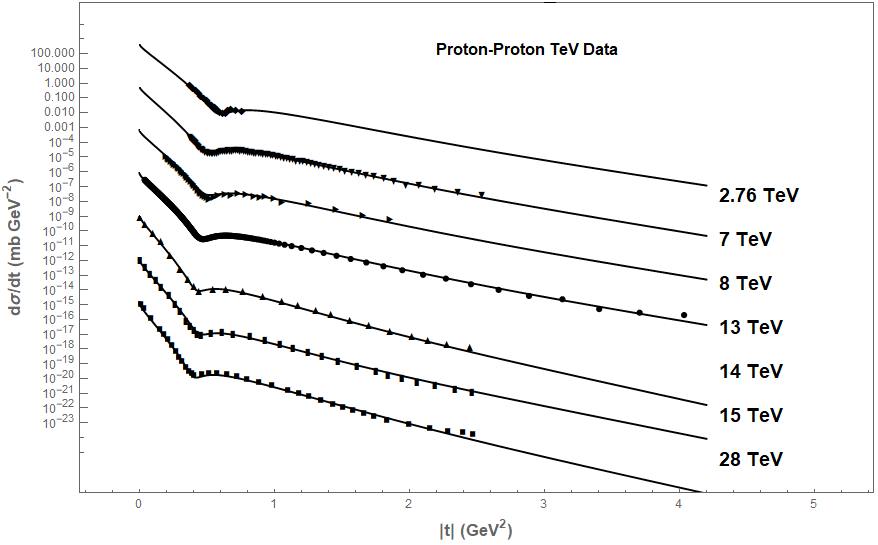}
			\caption{}
			\label{fig:a}
		\end{subfigure}
		\hfill
		\begin{subfigure}[b]{0.49\textwidth}
			\centering
			\includegraphics[width=\textwidth]{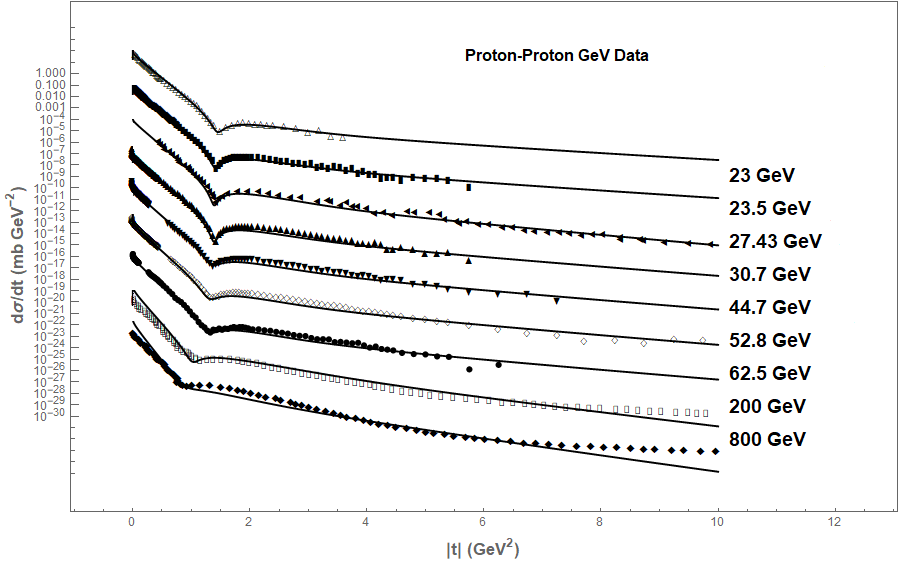}
			\caption{}
			\label{fig:b}
		\end{subfigure}
		\caption{(a) Same legend as in Fig.1 with Model 5. (b) Same legend as in Fig.1 with Model 5.}
		\label{fig:combined}
	\end{figure}
	\begin{figure}[h!]
		\centering
		\begin{subfigure}[b]{0.49\textwidth}
			\centering
			\includegraphics[width=\textwidth]{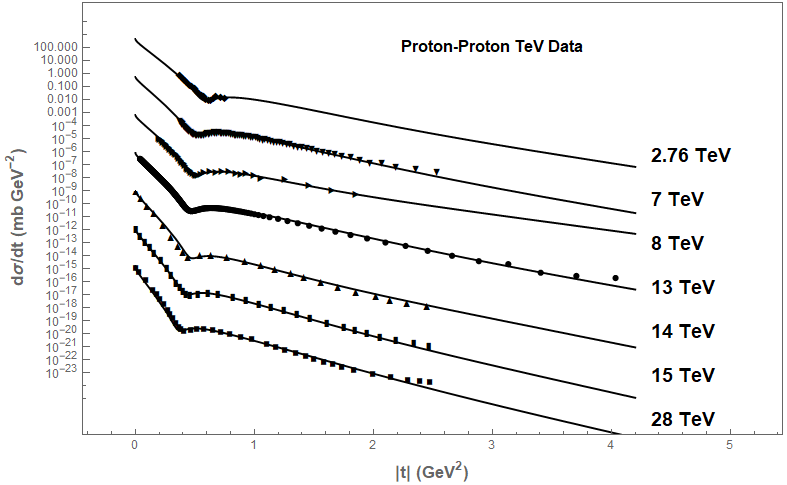}
			\caption{}
			\label{fig:a}
		\end{subfigure}
		\hfill
		\begin{subfigure}[b]{0.49\textwidth}
			\centering
			\includegraphics[width=\textwidth]{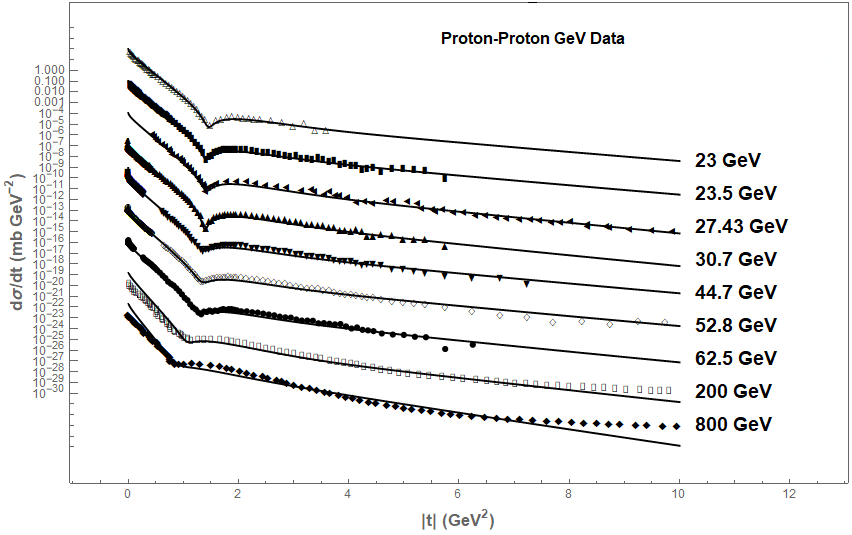}
			\caption{}
			\label{fig:b}
		\end{subfigure}
		\caption{(a) Same legend as in Fig.1 with Model 6. (b) Same legend as in Fig.1 with Model 6.}
		\label{fig:combined}
	\end{figure}
		\begin{figure}[h!]
		\centering
		\begin{subfigure}[b]{0.49\textwidth}
			\centering
			\includegraphics[width=\textwidth]{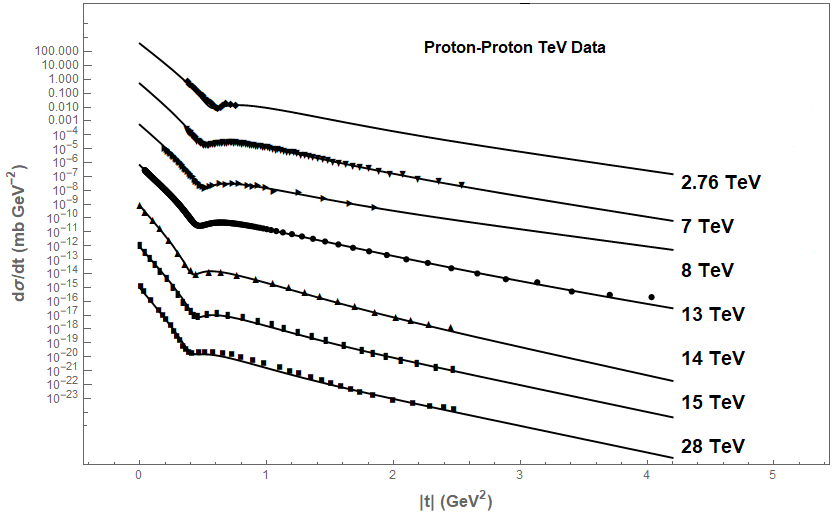}
			\caption{}
			\label{fig:a}
		\end{subfigure}
		\hfill
		\begin{subfigure}[b]{0.49\textwidth}
			\centering
			\includegraphics[width=\textwidth]{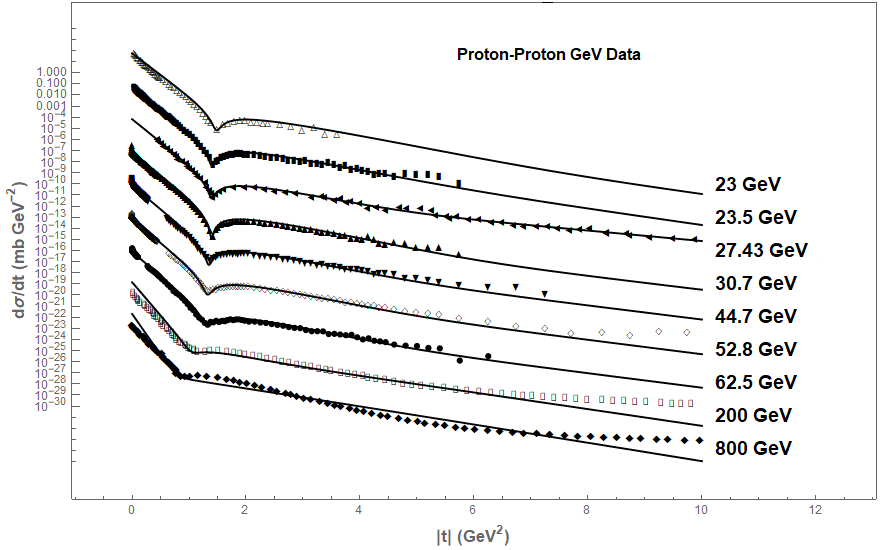}
			\caption{}
			\label{fig:b}
		\end{subfigure}
		\caption{(a) Same legend as in Fig.1 with Model 7. (b) Same legend as in Fig.1 with Model 7.}
		\label{fig:combined}
	\end{figure}
	\renewcommand{\arraystretch}{1.5}
	\setlength{\tabcolsep}{8pt}
	\begin{table}[h]
		\centering
		\caption{Parameter values of the Model 1.}
		\label{tab:parameters}
		\scalebox{0.60}{
			\begin{tabular}{c|ccccccccccc}
				\toprule
				\textbf{${\boldsymbol{\sqrt{s}}}$} & $\boldsymbol{A_1}$ & $\boldsymbol{A_2}$ & $\boldsymbol{A_3}$ & $\boldsymbol{A_4}$ & $\boldsymbol{B_1}$ & $\boldsymbol{B_2}$ & $\boldsymbol{B_3}$ & $\boldsymbol{B_4}$ & $\boldsymbol{\eta}$ & $\boldsymbol{\mu}$ & $\boldsymbol{\chi^{2}}$ \\
				\midrule
				& ($mb/\textrm{GeV}^{2}$) & ($mb/\textrm{GeV}^{2}$) & ($mb/\textrm{GeV}^{2}$) & ($mb/\textrm{GeV}^{2}$) & ($\textrm{GeV}^{-2}$) & ($\textrm{GeV}^{-2}$) & ($\textrm{GeV}^{-2}$) & ($\textrm{GeV}^{-2}$) &  &  & \\
				\midrule
				23 ${\textrm{GeV}}$ & 20.7284 & 5$\times10^{-5}$ & -0.06 & 0.04 & 9.20 & 1.59 & 3.00 & 2.754 & 0.17 & 1$\times10^{-5}$ & 0.03080 \\
				23.5 ${\textrm{GeV}}$ & 24.3006 & 1$\times10^{-4}$ & -0.06 & 0.04 & 9.44 & 1.50 & 3.00 & 2.754 & 0.1506 & 2$\times10^{-4}$ & 0.0130 \\
				27.43 ${\textrm{GeV}}$ & 25.9808 & 1$\times10^{-
					4}$ & -0.06 & 0.04 & 9.57 & 1.29 & 3.00 & 2.754 & 0.1503 & 0.004 & 0.0486 \\
				30.7 ${\textrm{GeV}}$ & 25.944 & 1$\times10^{-4}$ & -0.06 & 0.04 & 9.48 & 1.59 & 3.00 & 2.754 & 0.1503 & 0.005 & 0.5260 \\
				44.7 ${\textrm{GeV}}$ & 22.492 & 3$\times10^{-4}$ & -0.06 & 0.04 & 9.65 & 1.59 & 3.00 & 2.754 & 0.152 & 0.003 & 0.5809 \\
				52.8 ${\textrm{GeV}}$ & 22.1529 & 3$\times10^{-4}$ & -0.06 & 0.04 & 9.67 & 1.59 & 3.00 & 2.754 & 0.1503 & 0.005 & 0.4373 \\
				62.5 ${\textrm{GeV}}$ & 22.000 & 3$\times10^{-4}$ & -0.06 & 0.04 & 9.67 & 1.59 & 3.00 & 2.754 & 0.1503 & 0.005 & 0.2688 \\
				200 ${\textrm{GeV}}$ & 102.68321 & 1$\times10^{-4}$ & -0.06 & 0.01 & 8.65 & 1.00 & 3.00 & 2.00 & 0.025 & 0.04 & 36.9703 \\
				800 ${\textrm{GeV}}$ & 120.68321 & 1$\times10^{-4}$ & -0.06 & 0.01 & 8.50 & 1.00 & 3.00 & 2.00 & 0.038 & 0.05 & 85.2839 \\
				2.76 ${\textrm{TeV}}$ & 234.327 & 1.305 & -9.06 & 0.04 & 15.15 & 4.40 & 7.207 & 3.10 & 0.200 & 1$\times10^{-4}$ & 0.0132 \\
				7 ${\textrm{TeV}}$ & 292.000 & 1.305 & -9.06 & 0.04 & 15.50 & 4.30 & 7.207 & 3.10 & 0.103 & 0.034 & 0.0035\\
				8 ${\textrm{TeV}}$ & 320 & 1.305 & -9.06 & 0.04 & 15.95 & 4.20 & 7.207 & 2.925 & 0.105 & 0.025 & 0.0027 \\
				13 ${\textrm{TeV}}$ & 375.432 & 1.305 & -9.06 & 0.04 & 16.50 & 4.10 & 7.207 & 2.894 & 0.103 & 0.034 & 0.0003 \\
				14 ${\textrm{TeV}}$ & 670.487 & 1.305 & -9.06 & 0.04 & 16.20 & 4.60 & 7.207 & 3.00 & 0.04 & 0.08 & 0.0282 \\
				15 ${\textrm{TeV}}$ & 675.758 & 1.305 & -9.06 & 0.04 & 16.00 & 4.60 & 7.207 & 2.894 & 0.035 & 0.07 & 0.0927 \\
				28 ${\textrm{TeV}}$ & 695.734 & 1.305 & -9.06 & 0.04 & 16.70 & 4.40 & 7.207 & 2.894 & 0.05 & 0.07 & 0.0635 \\
				\bottomrule
			\end{tabular}
		}
	\end{table}
	\renewcommand{\arraystretch}{1.5}
	\setlength{\tabcolsep}{8pt}
	\begin{table}[h]
		\centering
		\caption{Parameter values of the Model 2.}
		\label{tab:parameters}
		\scalebox{0.60}{
			\begin{tabular}{c|cccccccccccccc}
				\toprule
				\textbf{${\boldsymbol{\sqrt{s}}}$} & $\boldsymbol{A_1}$ & $\boldsymbol{A_2}$ & $\boldsymbol{A_3}$ & $\boldsymbol{A_4}$ & $\boldsymbol{B_1}$ & $\boldsymbol{C_1}$ & $\boldsymbol{B_2}$ & $\boldsymbol{B_3}$ & $\boldsymbol{B_4}$ & $\boldsymbol{\alpha}$ & $\boldsymbol{m}$ & $\boldsymbol{\eta}$ & $\boldsymbol{\mu}$ & $\boldsymbol{\chi^{2}}$ \\
				\midrule
				& ($mb/\textrm{GeV}^{2}$) & ($mb/\textrm{GeV}^{2}$) & ($mb/\textrm{GeV}^{2}$) & ($mb/\textrm{GeV}^{2(1-m)}$) & ($\textrm{GeV}^{-2}$) & ($\textrm{GeV}^{-2}$) & ($\textrm{GeV}^{-2}$) & ($\textrm{GeV}^{-2}$) & ($\textrm{GeV}^{-2}$) &  &  &  &  & \\
				\midrule
				23 ${\textrm{GeV}}$ & 20.420 & 1$\times10^{-5}$ & -0.06 & 0.02 & 7.92 & 5.00 & 0.83 & 3.00 & 2.374 & 0.18 & 0.1 & 0.145 & 0.018 & 0.0387 \\
				23.5 ${\textrm{GeV}}$ & 22.022 & 1$\times10^{-5}$ & -0.06 & 0.02 & 8.00 & 5.00 & 0.83 & 3.00 & 2.374 & 0.18 & 0.1 & 0.147 & 0.023 & 0.0096 \\
				27.43 ${\textrm{GeV}}$ & 24.751 & 1$\times10^{-5}$ & -0.06 & 0.02 & 8.08 & 5.00 & 0.83 & 3.00 & 2.374 & 0.18 & 0.1 & 0.146 & 0.022 & 0.0179 \\
				30.7 ${\textrm{GeV}}$ & 22.200 & 1$\times10^{-5}$ & -0.06 & 0.02 & 8.03 & 5.00 & 0.83 & 3.00 & 2.374 & 0.18 & 0.1 & 0.146 & 0.022 & 0.5338 \\
				44.7 ${\textrm{GeV}}$ & 22.264 & 1$\times10^{-5}$ & -0.06 & 0.02 & 8.00 & 5.00 & 0.82 & 3.00 & 2.374 & 0.18 & 0.1 & 0.147 & 0.027 & 0.5365 \\
				52.8 ${\textrm{GeV}}$ & 20.645 & 1$\times10^{-5}$ & -0.06 & 0.02 & 8.00 & 5.00 & 0.83 & 3.00 & 2.374 & 0.18 & 0.1 & 0.147 & 0.027 & 0.3684 \\
				62.5 ${\textrm{GeV}}$ & 20.694 & 1$\times10^{-5}$ & -0.06 & 0.02 & 8.00 & 5.00 & 0.83 & 3.00 & 2.374 & 0.18 & 0.1 & 0.147 & 0.027 & 0.1901 \\
				200 ${\textrm{GeV}}$ & 108.683 & 0.01 & -0.06 & 0.01 & 10.90 & 5.50 & 3.70 & 3.00 & 1.60 & 0.18 & 0.8 & 0.008 & 0.018 & 38.7873 \\
				800 ${\textrm{GeV}}$ & 132.683 & 0.01 & -0.06 & 0.02 & 10.0 & 5.00 & 3.00 & 3.00 & 1.60 & 0.18 & 0.8 & 0.008 & 0.024 & 66.1178 \\
				2.76 ${\textrm{TeV}}$ & 220.870 & 1.20 & -9.06 & 0.04 & 15.250 & 5.80 & 4.35 & 7.207 & 2.454 & 0.2 & 1.00 & 0.150 & 0.001 & 0.0860 \\
				7 ${\textrm{TeV}}$ & 225.000 & 1.20 & -9.06 & 0.04 & 15.02 & 4.00 & 4.45 & 7.207 & 2.454 & 0.2 & 1.00 & 0.1503 & 0.042 & 0.0038 \\
				8 ${\textrm{TeV}}$ & 162.996 & 1.10 & -9.06 & 0.04 & 15.25 & 5.70 & 4.40 & 7.50 & 2.55 & 0.5 & 1.00 & 0.17 & 0.015 & 0.0002 \\
				13 ${\textrm{TeV}}$ & 251.513 & 1.20 & -9.06 & 0.04 & 15.272 & 4.00 & 4.35 & 7.207 & 2.454 & 0.2 & 1.00 & 0.1503 & 0.045 & 0.0001 \\
				14 ${\textrm{TeV}}$ & 510.123 & 1.20 & -9.06 & 0.04 & 15.70 & 4.00 & 4.70 & 7.207 & 2.654 & 0.2 & 1.00 & 0.043 & 0.07 & 0.0454 \\
				15 ${\textrm{TeV}}$ & 535.025 & 1.20 & -9.06 & 0.04 & 16.0 & 4.00 & 4.60 & 7.207 & 2.954 & 0.2 & 1.00 & 0.04 & 0.06 & 0.0978 \\
				28 ${\textrm{TeV}}$ & 560.250 & 1.20 & -9.06 & 0.04 & 16.50 & 5.00 & 4.55 & 7.207 & 2.654 & 0.2 & 1.00 & 0.0503 & 0.06 & 0.0683 \\
				\bottomrule
			\end{tabular}
		}
	\end{table}
	\renewcommand{\arraystretch}{1.5}
	\setlength{\tabcolsep}{8pt}
	\begin{table}[h]
		\centering
		\caption{Parameter values of the Model 3.}
		\label{tab:parameters}
		\scalebox{0.60}{
			\begin{tabular}{c|ccccccccccccc}
				\toprule
				\textbf{${\boldsymbol{\sqrt{s}}}$} & $\boldsymbol{A_1}$ & $\boldsymbol{A_2}$ & $\boldsymbol{A_3}$ & $\boldsymbol{A_4}$ & $\boldsymbol{B_1}$ & $\boldsymbol{B_2}$ & $\boldsymbol{B_3}$ & $\boldsymbol{B_4}$ & $\boldsymbol{\alpha}$ & $\boldsymbol{m}$ & $\boldsymbol{\eta}$ & $\boldsymbol{\mu}$ & $\boldsymbol{\chi^{2}}$ \\
				\midrule
				& ($mb/\textrm{GeV}^{2}$) & ($mb/\textrm{GeV}^{2}$) & ($mb/\textrm{GeV}^{2}$) & ($mb/\textrm{GeV}^{2(1-m)}$) & ($\textrm{GeV}^{-2}$) & ($\textrm{GeV}^{-2}$) & ($\textrm{GeV}^{-2}$) & ($\textrm{GeV}^{-2}$) &  & &  &  & \\
				\midrule
				23 ${\textrm{GeV}}$ & 24.000 & 1$\times10^{-5}$ & -0.06 & 0.02 & 8.03 & 0.53 & 3.00 & 2.374 & 0.1 & 0.1 & 0.145 & 0.018 & 0.0311 \\
				23.5 ${\textrm{GeV}}$ & 24.618 & 1$\times10^{-5}$ & -0.06 & 0.02 & 8.075 & 0.53 & 3.00 & 2.374 & 0.1 & 0.1 & 0.147 & 0.023 & 0.0140 \\
				27.43 ${\textrm{GeV}}$ & 26.5304 & 1$\times10^{-5}$ & -0.06 & 0.02 & 8.13 & 0.53 & 3.00 & 2.374 & 0.1 & 0.1 & 0.146 & 0.023 & 0.0176 \\
				30.7 ${\textrm{GeV}}$ & 24.805 & 1$\times10^{-5}$ & -0.06 & 0.02 & 8.11 & 0.53 & 3.00 & 2.374 & 0.1 & 0.1 & 0.146 & 0.023 & 0.5542 \\
				44.7 ${\textrm{GeV}}$ & 23.682 & 1$\times10^{-5}$ & -0.06 & 0.02 & 8.00 & 0.53 & 3.00 & 2.374 & 0.1 & 0.1 & 0.147 & 0.027 & 0.5759 \\
				52.8 ${\textrm{GeV}}$ & 22.539 & 1$\times10^{-5}$ & -0.06 & 0.02 & 8.00 & 0.53 & 3.00 & 2.374 & 0.1 & 0.1 & 0.147 & 0.027 & 0.4412 \\
				62.5 ${\textrm{GeV}}$ & 23.0986 & 1$\times10^{-5}$ & -0.06 & 0.02 & 8.00 & 0.53 & 3.00 & 2.374 & 0.1 & 0.1 & 0.147 & 0.027 & 0.2581 \\
				200 ${\textrm{GeV}}$ & 108.683 & 0.01 & -0.06 & 0.01 & 10.90 & 3.70 & 3.00 & 1.60 & 0.1 & 0.8 & 0.018 & 0.018 & 35.7519 \\
				800 ${\textrm{GeV}}$ & 120.000 & 0.001 & -0.06 & 0.01 & 8.50 & 0.90 & 3.00 & 2.00 & 0.1 & 0.8 & 0.038 & 0.05 & 85.0032 \\
				2.76 ${\textrm{TeV}}$ & 225.036 & 1.20 & -9.06 & 0.04 & 14.000 & 4.350 & 7.207 & 2.454 & 0.1 & 1.0 & 0.150 & 0.001 & 0.0010 \\
				7 ${\textrm{TeV}}$ & 250.000 & 1.20 & -9.06 & 0.04 & 15.10 & 4.10 & 7.207 & 2.754 & 0.1 & 1.0 & 0.1503 & 0.04 & 0.0475 \\
				8 ${\textrm{TeV}}$ & 270.209 & 1.20 & -9.06 & 0.04 & 16.40 & 4.30 & 7.50 & 2.55 & 0.1 & 1.0 & 0.1603 & 0.028 & 0.0143 \\
				13 ${\textrm{TeV}}$ & 299.1 & 1.20 & -9.06 & 0.04 & 15.733 & 4.160 & 7.207 & 2.454 & 0.1 & 1.0 & 0.1503 & 0.045 & 0.0001 \\
				14 ${\textrm{TeV}}$ & 400.958 & 1.200 & -9.06 & 0.04 & 16.0 & 4.70 & 7.50 & 2.90 & 0.1 & 1.0 & 0.020 & 0.070 & 0.2922 \\
				15 ${\textrm{TeV}}$ & 415.373 & 1.200 & -9.06 & 0.04 & 16.20 & 4.70 & 7.50 & 2.80 & 0.1 & 1.0 & 0.02 & 0.062 & 0.3553 \\
				28 ${\textrm{TeV}}$ & 490.411 & 1.20 & -9.06 & 0.04 & 17.10 & 4.20 & 7.50 & 2.80 & 0.1 & 1.0 & 0.022 & 0.065 & 0.3120 \\
				\bottomrule
			\end{tabular}
		}
	\end{table}
	\renewcommand{\arraystretch}{1.5}
	\setlength{\tabcolsep}{8pt}
	\begin{table}[h]
		\centering
		\caption{Parameter values of the Model 4.}
		\label{tab:parameters}
		\scalebox{0.60}{
			\begin{tabular}{c|cccccccccccccc}
				\toprule
				\textbf{${\boldsymbol{\sqrt{s}}}$} & $\boldsymbol{A_1}$ & $\boldsymbol{A_2}$ & $\boldsymbol{A_3}$ & $\boldsymbol{A_4}$ & $\boldsymbol{B_1}$ & $\boldsymbol{B_2}$ & $\boldsymbol{B_3}$ & $\boldsymbol{B_4}$ & $\boldsymbol{\alpha}$ & $\boldsymbol{\beta}$ & $\boldsymbol{t_0}$ & $\boldsymbol{\eta}$ & $\boldsymbol{\mu}$ & $\boldsymbol{\chi^{2}}$ \\
				\midrule
				& ($mb/\textrm{GeV}^{2}$) & ($mb/\textrm{GeV}^{2}$) & ($mb/\textrm{GeV}^{2}$) & ($mb/\textrm{GeV}^{2}$) & ($\textrm{GeV}^{-2}$) & ($\textrm{GeV}^{-2}$) & ($\textrm{GeV}^{-2}$) & ($\textrm{GeV}^{-2}$) &  &  & ($\textrm{GeV}^{2}$) &  &  & \\
				\midrule
				23 ${\textrm{GeV}}$ & 24.946 & 1$\times10^{-5}$ & -0.06 & 0.02 & 8.12 & 0.90 & 3.00 & 2.364 & 0.1 & 0.07 & 1.0 & 0.136 & 0.007 & 0.0323 \\
				23.5 ${\textrm{GeV}}$ & 24.316 & 1$\times10^{-5}$ & -0.06 & 0.02 & 8.125 & 0.68 & 3.00 & 2.364 & 0.1 & 0.08 & 1.0 & 0.146 & 0.015 & 0.0162 \\
				27.43 ${\textrm{GeV}}$ & 24.315 & 1$\times10^{-5}$ & -0.06 & 0.02 & 8.05 & 0.68 & 3.00 & 2.364 & 0.1 & 0.07 & 1.0 & 0.146 & 0.014 & 0.0199 \\
				30.7 ${\textrm{GeV}}$ & 23.646 & 1$\times10^{-5}$ & -0.06 & 0.02 & 8.12 & 0.68 & 3.00 & 2.364 & 0.1 & 0.02 & 1.0 & 0.146 & 0.014 & 0.5720 \\
				44.7 ${\textrm{GeV}}$ & 23.289 & 1$\times10^{-5}$ & -0.06 & 0.02 & 8.00 & 0.72 & 3.00 & 2.372 & 0.1 & 0.02 & 1.0 & 0.1402 & 0.020 & 0.6005 \\
				52.8 ${\textrm{GeV}}$ & 22.817 & 1$\times10^{-5}$ & -0.06 & 0.02 & 8.00 & 0.68 & 3.00 & 2.364 & 0.1 & 0.02 & 1.0 & 0.1407 & 0.018 & 0.4579 \\
				62.5 ${\textrm{GeV}}$ & 23.324 & 1$\times10^{-5}$ & -0.06 & 0.02 & 8.0 & 0.68 & 3.0 & 2.364 & 0.1 & 0.02 & 1.0 & 0.1407 & 0.020 & 0.2792\\
				200 ${\textrm{GeV}}$ & 101.683 & 0.01 & -0.06 & 0.01 & 9.80 & 3.70 & 3.00 & 1.80 & 0.1 & 0.002 & 1.0 & 0.018 & 0.018 & 32.5794 \\
				800 ${\textrm{GeV}}$ & 122.000 & 0.001 & -0.06 & 0.01 & 8.50 & 0.950 & 3.00 & 2.00 & 0.1 & 0.002 & 1.0 & 0.038 & 0.05 & 86.8693 \\
				2.76 ${\textrm{TeV}}$ & 225.316 & 1.20 & -9.06 & 0.04 & 15.210 & 4.05 & 7.407 & 2.454 & 0.1 & 0.2 & 1.0 & 0.250 & 3$\times10^{-4}$ & 0.0438 \\
				7 ${\textrm{TeV}}$ & 235.999 & 1.20 & -9.06 & 0.04 & 14.40 & 4.075 & 7.207 & 2.60 & 0.1 & 0.2 & 1.0 & 0.1703 & 0.05 & 0.0153 \\
				8 ${\textrm{TeV}}$ & 262.409 & 1.20 & -9.06 & 0.04 & 15.85 & 4.00 & 7.50 & 2.75 & 0.1 & 0.2 & 1.0 & 0.1503 & 0.02 & 0.0007 \\
				13 ${\textrm{TeV}}$ & 313.521 & 1.20 & -9.06 & 0.04 & 15.70 & 3.75 & 7.207 & 2.654 & 0.1 & 0.2 & 1.0 & 0.1503 & 0.05 & $\approx0$ \\
				14 ${\textrm{TeV}}$ & 525.994 & 1.20 & -9.06 & 0.04 & 14.82 & 4.450 & 7.207 & 3.10 & 0.1 & 0.2 & 1.0 & 0.074 & 0.084 & 0.0445 \\
				15 ${\textrm{TeV}}$ & 550.376 & 1.20 & -9.06 & 0.04 & 14.85 & 4.450 & 7.207 & 3.10 & 0.1 & 0.2 & 1.0 & 0.075 & 0.08 & 0.0885 \\
				28 ${\textrm{TeV}}$ & 570.456 & 1.20 & -9.06 & 0.04 & 15.30 & 4.250 & 7.207 & 2.684 & 0.1 & 0.2 & 1.0 & 0.08 & 0.08 & 0.0619 \\
				\bottomrule
			\end{tabular}
		}
	\end{table}
	\renewcommand{\arraystretch}{1.5}
	\setlength{\tabcolsep}{8pt}
	\begin{table}[h]
		\centering
		\caption{Parameter values of the Model 5.}
		\label{tab:parameters}
		\scalebox{0.58}{
			\begin{tabular}{c|ccccccccccccccc}
				\toprule
				\textbf{${\boldsymbol{\sqrt{s}}}$} & $\boldsymbol{A_1}$ & $\boldsymbol{A_2}$ & $\boldsymbol{A_3}$ & $\boldsymbol{A_4}$ & $\boldsymbol{B_1}$ & $\boldsymbol{B_2}$ & $\boldsymbol{B_3}$ & $\boldsymbol{B_4}$ & $\boldsymbol{m}$ & $\boldsymbol{n}$ & $\boldsymbol{p}$ & $\boldsymbol{t_0}$ & $\boldsymbol{\eta}$ & $\boldsymbol{\mu}$ & $\boldsymbol{\chi^{2}}$ \\
				\midrule
				& ($mb/\textrm{GeV}^{2(1-m)}$) & ($mb/\textrm{GeV}^{2(1-n)}$) & ($mb/\textrm{GeV}^{2}$) & ($mb/\textrm{GeV}^{2}$) & ($\textrm{GeV}^{-2}$) & ($\textrm{GeV}^{-2}$) & ($\textrm{GeV}^{-2}$) & ($\textrm{GeV}^{-2}$) &  &  &  & ($\textrm{GeV}^{2}$) &  &  & \\
				\midrule
				23 ${\textrm{GeV}}$ & 24.939 & 1$\times10^{-4}$ & -0.06 & 0.01 & 7.305 & 0.45 & 3.0 & 2.2 & 0.15 & 1.30 & 0.3 & 10 & 0.06 & 0.034 & 0.0412\\
				23.5 ${\textrm{GeV}}$ & 24.309 & 1$\times10^{-4}$ & -0.06 & 0.01 & 7.32 & 0.49 & 3.0 & 2.19 & 0.2 & 1.34 & 0.3 & 10 & 0.05 & 0.037 & 0.0127 \\
				27.43 ${\textrm{GeV}}$ & 32.7578 & 1$\times10^{-4}$ & -0.06 & 0.01 & 7.40 & 0.69 & 3.00 & 2.20 & 0.10 & 1.30 & 0.3 & 10 & 0.08 & 0.042 & 0.0972 \\
				30.7 ${\textrm{GeV}}$ & 40.2425 & 1$\times10^{-4}$ & -0.06 & 0.01 & 7.475 & 0.59 & 3.0 & 2.216 & 0.20 & 1.30 & 0.3 & 10 & 0.005 & 0.042 & 0.1691 \\
				44.7 ${\textrm{GeV}}$ & 28.0184 & 1$\times10^{-4}$ & -0.06 & 0.01 & 7.26 & 0.59 & 3.0 & 2.190 & 0.20 & 1.34 & 0.3 & 10 & 0.05 & 0.04 & 0.2123 \\
				52.8 ${\textrm{GeV}}$ & 26.5816 & 1$\times10^{-4}$ & -0.06 & 0.01 & 7.23 & 0.59 & 3.0 & 2.190 & 0.20 & 1.34 & 0.3 & 10 & 0.05 & 0.042 & 0.0964 \\
				62.5 ${\textrm{GeV}}$ & 25.7203 & 1$\times10^{-4}$ & -0.06 & 0.01 & 7.22 & 0.59 & 3.0 & 2.190 & 0.20 & 1.34 & 0.3 & 10 & 0.05 & 0.042 & 0.0180 \\
				200 ${\textrm{GeV}}$ & 67.000 & 0.001 & -0.06 & 0.01 & 9.70 & 0.97 & 3.0 & 2.0 & 0.20 & 1.34 & 0.3 & 10 & 0.0005 & 0.026 & 42.3886 \\
				800 ${\textrm{GeV}}$ & 87.000 & 0.001 & -0.06 & 0.01 & 8.40 & 0.88 & 3.00 & 2.00 & 0.20 & 1.34 & 0.3 & 10 & 0.0005 & 0.04 & 116.463 \\
				2.76 ${\textrm{TeV}}$ & 180.000 & 0.99 & -9.06 & 0.04 & 13.67 & 4.2 & 7.07 & 2.854 & 0.1 & 0.01 & 0.1 & 20 & 0.15 & 0.043 & 0.0566 \\
				7 ${\textrm{TeV}}$ & 175.000 & 0.99 & -9.06 & 0.04 & 14.30 & 4.2 & 7.307 & 2.854 & 0.1 & 0.01 & 0.1 & 20 & 0.15 & 0.045 & 0.0040 \\
				8 ${\textrm{TeV}}$ & 187.000 & 1.3 & -9.06 & 0.04 & 14.20 & 4.05 & 6.77 & 2.754 & 0.1 & 0.01 & 0.1 & 20 & 0.15 & 0.050 & 0.0332 \\
				13 ${\textrm{TeV}}$ & 195.670 & 1.3 & -9.06 & 0.04 & 14.4527 & 4.0 & 6.82 & 2.754 & 0.1 & 0.01 & 0.1 & 20 & 0.15 & 0.045 & 0.0032 \\
				14 ${\textrm{TeV}}$ & 480.056 & 1.0 & -9.06 & 0.04 & 19.40 & 5.40 & 9.10 & 3.82 & 0.1 & 0.01 & 0.1 & 20 & 0.012 & 0.036 & 0.0245 \\
				15 ${\textrm{TeV}}$ & 510.723 & 1.1 & -9.04 & 0.04 & 21.00 & 6.0 & 9.80 & 3.9 & 0.1 & 0.01 & 0.1 & 20 & 0.12 & 0.015 & 0.0577 \\
				28 ${\textrm{TeV}}$ & 550.817 & 1.1 & -9.04 & 0.04 & 22.10 & 5.70 & 9.80 & 4.200 & 0.1 & 0.01 & 0.1 & 20 & 0.02 & 0.015 & 0.0569 \\
				\bottomrule
			\end{tabular}
		}
	\end{table}
	\renewcommand{\arraystretch}{1.5}
	\setlength{\tabcolsep}{8pt}
	\begin{table}[h]
		\centering
		\caption{Parameter values of the Model 6.}
		\label{tab:parameters}
		\scalebox{0.58}{
			\begin{tabular}{c|ccccccccccccccc}
				\toprule
				\textbf{${\boldsymbol{\sqrt{s}}}$} & $\boldsymbol{A_1}$ & $\boldsymbol{A_2}$ & $\boldsymbol{A_3}$ & $\boldsymbol{A_4}$ & $\boldsymbol{B_1}$ & $\boldsymbol{B_2}$ & $\boldsymbol{B_3}$ & $\boldsymbol{B_4}$ & $\boldsymbol{m}$ & $\boldsymbol{n}$ & $\boldsymbol{\alpha}$ & $\boldsymbol{t_0}$ & $\boldsymbol{\eta}$ & $\boldsymbol{\mu}$ & $\boldsymbol{\chi^{2}}$ \\
				\midrule
				& ($mb/\textrm{GeV}^{2(1-m)}$) & ($mb/\textrm{GeV}^{2(1-n)}$) & ($mb/\textrm{GeV}^{2}$) & ($mb/\textrm{GeV}^{2}$) & ($\textrm{GeV}^{-2}$) & ($\textrm{GeV}^{-2}$) & ($\textrm{GeV}^{-2}$) & ($\textrm{GeV}^{-2}$) &  &  &  & ($\textrm{GeV}^{2}$) &  &  &  \\
				\midrule
				23 ${\textrm{GeV}}$ & 23.215 & 1$\times10^{-4}$ & -0.06 & 0.01 & 7.172 & 0.69 & 3.00 & 2.190 & 0.20 & 1.34 & 2.7 & 1.0 & 0.05 & 0.035 & 0.0336 \\
				23.5 ${\textrm{GeV}}$ & 24.4095 & 1$\times10^{-4}$ & -0.06 & 0.01 & 7.210 & 0.69 & 3.00 & 2.190 & 0.20 & 1.34 & 2.7 & 1.0 & 0.05 & 0.04 & 0.0123 \\
				27.43 ${\textrm{GeV}}$ & 28.014 & 1$\times10^{-4}$ & -0.06 & 0.01 & 7.210 & 0.79 & 3.00 & 2.190 & 0.20 & 1.34 & 2.7 & 1.0 & 0.05 & 0.04 & 0.1449 \\
				30.7 ${\textrm{GeV}}$ & 29.369 & 1$\times10^{-4}$ & -0.06 & 0.01 & 7.30 & 0.79 & 3.00 & 2.190 & 0.20 & 1.34 & 2.7 & 1.0 & 0.05 & 0.04 & 0.1738 \\
				44.7 ${\textrm{GeV}}$ & 28.0497 & 1$\times10^{-4}$ & -0.06 & 0.01 & 7.20 & 0.69 & 3.00 & 2.190 & 0.20 & 1.34 & 2.7 & 1.0 & 0.05 & 0.04 & 0.2160 \\
				52.8 ${\textrm{GeV}}$ & 26.5437 & 1$\times10^{-4}$ & -0.06 & 0.01 & 7.15 & 0.69 & 3.00 & 2.190 & 0.20 & 1.34 & 2.7 & 1.0 & 0.05 & 0.04 & 0.0999 \\
				62.5 ${\textrm{GeV}}$ & 25.736 & 1$\times10^{-4}$ & -0.06 & 0.01 & 6.70 & 0.60 & 3.00 & 2.290 & 0.20 & 1.34 & 2.7 & 1.0 & 0.05 & 0.05 & 0.0192 \\
				200 ${\textrm{GeV}}$ & 60.000 & 0.0001 & -0.06 & 0.01 & 8.22 & 0.76 & 3.00 & 2.00 & 0.20 & 1.34 & 2.40 & 1.0 & 0.01 & 0.04 & 41.8613 \\
				800 ${\textrm{GeV}}$ & 75.000 & 0.001 & -0.06 & 0.01 & 8.22 & 0.96 & 3.0 & 2.0 & 0.20 & 1.34 & 2.4 & 1.0 & 0.01 & 0.04 & 111.098 \\
				2.76 ${\textrm{TeV}}$ & 180.341 & 1.30 & -9.06 & 0.04 & 13.620 & 4.75 & 7.205 & 2.954 & 0.1 & 0.01 & 2.7 & 1.0 & 0.1503 & 0.045 & 0.0741 \\
				7 ${\textrm{TeV}}$ & 185.545 & 1.30 & -9.06 & 0.04 & 13.90 & 4.45 & 7.205 & 2.954 & 0.1 & 0.01 & 2.7 & 1.0 & 0.1303 & 0.05 & 0.0042 \\
				8 ${\textrm{TeV}}$ & 187.374 & 1.30 & -9.06 & 0.04 & 14.00 & 4.58 & 7.255 & 2.354 & 0.1 & 0.01 & 2.7 & 1.0 & 0.1503 & 0.045 & 0.005 \\
				13 ${\textrm{TeV}}$ & 198.941 & 1.30 & -9.06 & 0.04 & 14.489 & 4.15 & 7.205 & 2.854 & 0.1 & 0.01 & 2.7 & 1.0 & 0.1503 & 0.045 & 0.0022 \\
				14 ${\textrm{TeV}}$ & 415.000 & 1.10 & -9.04 & 0.04 & 19.30 & 6.20 & 9.80 & 3.80 & 0.1 & 0.01 & 2.7 & 1.0 & 0.030 & 0.020 & 0.0338 \\
				15 ${\textrm{TeV}}$ & 475.146 & 1.10 & -9.04 & 0.04 & 19.90 & 5.80 & 9.80 & 4.00 & 0.1 & 0.01 & 2.7 & 1.0 & 0.031 & 0.030 & 0.0498 \\
				28 ${\textrm{TeV}}$ & 535.623 & 1.10 & -9.04 & 0.04 & 21.00 & 5.20 & 9.80 & 3.80 & 0.1 & 0.01 & 2.7 & 1.0 & 0.031 & 0.034 & 0.0538 \\
				\bottomrule
			\end{tabular}
		}
	\end{table}
		\renewcommand{\arraystretch}{1.5}
	\setlength{\tabcolsep}{8pt}
	\begin{table}[h]
		\centering
		\caption{Parameter values of the Model 7.}
		\label{tab:parameters}
		\scalebox{0.60}{
			\begin{tabular}{c|ccccccccccccccc}
				\toprule
				\textbf{${\boldsymbol{\sqrt{s}}}$} & $\boldsymbol{A_1}$ & $\boldsymbol{A_2}$ & $\boldsymbol{A_3}$ & $\boldsymbol{A_4}$ & $\boldsymbol{B_1}$ & $\boldsymbol{B_2}$ & $\boldsymbol{B_3}$ & $\boldsymbol{B_4}$ & $\boldsymbol{\gamma}$ & $\boldsymbol{m}$ & $\boldsymbol{t_0}$ & $\boldsymbol{n}$ & $\boldsymbol{\eta}$ & $\boldsymbol{\mu}$ & $\boldsymbol{\chi^{2}}$ \\
				\midrule
				& ($mb/\textrm{GeV}^{2}$) & ($mb/\textrm{GeV}^{2}$) & ($mb/\textrm{GeV}^{2}$) & ($mb/\textrm{GeV}^{2}$) & ($\textrm{GeV}^{-2}$) & ($\textrm{GeV}^{-2}$) & ($\textrm{GeV}^{-2}$) & ($\textrm{GeV}^{-2}$) &  &  & ($\textrm{GeV}^{2}$) &  &  &  &  \\
				\midrule
				23 ${\textrm{GeV}}$ & 22.240 & 3$\times10^{-5}$ & -0.06 & 0.04 & 9.170 & 1.79 & 3.00 & 2.754 & 0.001 & 1.1 & 20 & 5.5 & 0.1503 & 0.0002 & 0.0428 \\
				23.5 ${\textrm{GeV}}$ & 24.389 & 1$\times10^{-4}$ & -0.06 & 0.04 & 9.370 & 1.79 & 3.00 & 2.754 & 0.001 & 1.1 & 20 & 5.5 & 0.1503 & 0.006 & 0.0128 \\
				27.43 ${\textrm{GeV}}$ & 24.968 & 1$\times10^{-5}$ & -0.06 & 0.04 & 9.250 & 1.20 & 3.00 & 2.754 & 0.001 & 1.1 & 20 & 5.5 & 0.1503 & 0.009 & 0.0268 \\
				30.7 ${\textrm{GeV}}$ & 24.999 & 1$\times10^{-5}$ & -0.06 & 0.04 & 9.260 & 1.50 & 3.00 & 2.754 & 0.001 & 1.1 & 20 & 5.5 & 0.1503 & 0.009 & 0.5402 \\
				44.7 ${\textrm{GeV}}$ & 23.464 & 3$\times10^{-5}$ & -0.06 & 0.04 & 9.250 & 1.50 & 3.00 & 2.754 & 0.001 & 1.1 & 20 & 5.5 & 0.1503 & 0.012 & 0.5701 \\
				52.8 ${\textrm{GeV}}$ & 22.771 & 3$\times10^{-5}$ & -0.06 & 0.04 & 9.20 & 1.50 & 3.00 & 2.754 & 0.001 & 1.1 & 20 & 5.5 & 0.1503 & 0.014 & 0.4249 \\
				62.5 ${\textrm{GeV}}$ & 22.789 & 3$\times10^{-5}$ & -0.06 & 0.04 & 9.18 & 1.50 & 3.00 & 2.754 & 0.001 & 1.1 & 20 & 5.5 & 0.1503 & 0.014 & 0.2513 \\
				200 ${\textrm{GeV}}$ & 102.00 & 0.0003 & -0.06 & 0.04 & 10.20 & 1.40 & 3.00 & 2.754 & 0.001 & 1.1 & 20 & 5.5 & 0.032 & 0.03 & 41.9187 \\
				800 ${\textrm{GeV}}$ & 132.000 & 0.001 & -0.06 & 0.01 & 4.50 & 0.55 & 3.00 & 2.00 & 0.001 & 1.1 & 20 & 5.5 & 0.032 & 0.1 & 86.4296 \\
				2.76 ${\textrm{TeV}}$ & 293.629 & 1.30 & -9.70 & 0.04 & 14.82 & 4.68 & 7.35 & 2.79 & 0.001 & 1.1 & 20 & 0.1 & 0.1503 & 0.045 & 0.0970 \\
				7 ${\textrm{TeV}}$ & 290.745 & 1.30 & -9.70 & 0.04 & 15.0 & 4.4 & 7.23 & 2.79 & 0.001 & 1.1 & 20 & 0.1 & 0.1502 & 0.045 & 0.0415 \\
				8 ${\textrm{TeV}}$ & 297.261 & 1.30 & -9.06 & 0.04 & 15.05 & 4.3 & 7.0 & 2.35 & 0.001 & 1.1 & 20 & 0.1 & 0.1503 & 0.045 & 0.0001 \\
				13 ${\textrm{TeV}}$ & 498.282 & 1.30 & -9.06 & 0.04 & 16.91 & 3.80 & 7.00 & 2.754 & 0.001 & 1.1 & 20 & 0.1 & 0.06 & 0.044 & 0.0001 \\
				14 ${\textrm{TeV}}$ & 590.086 & 1.20 & -9.06 & 0.04 & 15.90 & 4.48 & 7.207 & 3.05 & 0.001 & 1.1 & 20 & 0.1 & 0.06 & 0.08 & 0.0339 \\
				15 ${\textrm{TeV}}$ & 605.075 & 1.20 & -9.06 & 0.04 & 15.30 & 4.70 & 7.207 & 2.884 & 0.001 & 1.1 & 20 & 0.1 & 0.06 & 0.08 & 0.0841 \\
				28 ${\textrm{TeV}}$ & 635.543 & 1.20 & -9.06 & 0.04 & 15.20 & 4.60 & 7.207 & 2.684 & 0.001 & 1.1 & 20 & 0.1 & 0.06 & 0.075 & 0.0627 \\
				\bottomrule
			\end{tabular}
		}
	\end{table}
	In most fits of our models, the chi-square per degree of freedom ($\chi^{2}$) values are reasonably small, which indicates good fitting agreement of our models with the data. At 200 GeV and 800 GeV, greater values of $\chi^{2}$ are observed due to the poor fit of our models at high t. We have used the following expression for dimensionless $\chi^{2}$.
	\begin{equation}
		\chi^2 = \frac{\sum_{i=1}^{N}[(d\sigma/dt)^{\text{model}}_i - (d\sigma/dt)^{\text{data}}_i]^{2}}{\sum_{i=1}^{N} [(d\sigma/dt)^{\text{data}}_i]^{2}}
	\end{equation}
	with i = 1, 2, 3,...,N, number of data points. It is a normalized error measure that is particularly useful to determine the goodness of fit. The $\chi^{2}$ values for each model across all the energies, along with the fitted parameter values, are given in Tables II-VIII. In some cases, especially for 200 GeV and 800 GeV data, the physical shape of the data and the optimal parameter values for the total elastic cross section are prioritized, which are critical for physical interpretation, even if chi-square values are large. The dimensionless \textsc{${\chi^{2}}$} is used as a relative indicator for comparing the performance of different models against the same dataset and allows possible identification of the most effective models. From Tables (II-VIII), it is observed that \textsc{${\chi^{2}}$} for the model 1 lies in range ($0.0003\leq{\chi^{2}}\leq85.2839$), for the model 2 lies in range ($0.0002\leq{\chi^{2}}\leq66.1178$), for the model 3 lies in range ($0.0001\leq{\chi^{2}}\leq85.0032$), for the model 4 lies in range ($0\leq{\chi^{2}}\leq86.8693$), for the model 5 lies in range $({0.0032\leq{\chi^{2}}\leq116.463})$, for the model 6 lies in range ($0.0022\leq{\chi^{2}}\leq111.098$), and for the model 7 lies in range ($0.0001\leq{\chi^{2}}\leq86.4296$). ${\chi^{2}}$ is reported equal to 0 for the values less than 0.0001. It is concluded that model 2 is best with respect to the least value of ${\chi^{2}}$. But, if we ignore the model fits at 200 GeV and 800 GeV, then ${\chi^{2}}$ is less than 0.61 for all the models, which show good agreement of these models with the experimental measured and extrapolated predicted available data.
	
	Our models contained fewer parameters as compared to the other earlier fits on the ISR and TOTEM data. A lesser number of parameters allows better modeling of the elastic differential cross section to obtain empirical information from the data. Fewer parameters reduce the chance of fitting noise or fluctuations in the data. Models with fewer parameters are very suitable in model-independent or phenomenological approaches, where the goal is to extract stable physical patterns from the experimental data. Such models are also used due to better stability when extrapolated beyond the range of current data to higher energies. With fewer parameter models, numerical fitting algorithms converge more quickly, computational efficiency is often improved, and estimation of uncertainties becomes straightforward.
	
	The proposed models for the differential cross section obtained through this work exhibit the same features as present in the data. At small ${\mid t\mid}$, the cross-section is dominated by a steep exponential fall-off. This region probes the large-distance or peripheral structure of the proton. As energy increases, the slope parameter B of the forward exponential becomes larger, a phenomenon known as the "shrinkage of the forward peak," as mentioned in refs. \cite{Antchev_2011} \cite{Antchev_2019}. The shrinkage is also explained by Regge phenomenology in ref. \cite{Collins:1977jy} and suggests that the proton becomes more transparent with an increase in energy, and its effective size increases. This indicates that the effective radius of interaction of the proton increases with center-of-mass energy. The shrinkage is often utilized to produce the slopes of effective trajectories in various elastic scattering phenomena. It is also important to study this phenomenon as it gives information about many observables, for instance, because of this phenomenon, the elastic cross-section becomes a decreasing fraction of the total cross-section as $log (s)\rightarrow\infty$.
	
	A prominent dip followed by a bump emerges at intermediate ${\mid t\mid}$ values around $\mid t\mid$ range (0.5–2.0 $\textrm{GeV}^{2}$), especially visible at the data at GeV energies and around $\mid t\mid$ range (0.1–0.4 $\textrm{GeV}^{2}$) at the data at TeV energies. The position of the dip shifts towards smaller ${\mid t\mid}$ values with increasing energy, indicating that the coherent interference between elastic amplitudes becomes more prominent at larger impact \cite{Antchev_2011}\cite{Amaldi1980}. At large ${\mid t\mid}$ values, the cross-section falls more slowly. This hard tail reflects deep elastic scattering between valence quarks, revealing the short-distance dynamics of the proton. Properly modeling this region requires either a modified exponential or a power-law behavior in the fit. Near the very small ${\mid t\mid}$ regime ($t\rightarrow0$), the cross-section tends to level off due to the finite proton size and effects of Coulomb-nuclear interference. Careful fitting in this region is crucial to accurately extract the total cross section via extrapolation. Our models adopt these features to varying degrees depending on how many exponential terms and corrections are included. The outer region composed of soft $q\bar{q}$ condensate cloud is probed at very small ${\mid t\mid}$ values. The middle region or baryonic shell influences the intermediate ${\mid t\mid}$ behavior where the dip appears. The core region composed of valence quarks is involved at large ${\mid t\mid}$ through hard scattering processes. Thus, the evolution of the shape of $\frac{d\sigma}{d|t|}$ with energy directly reflects the changing role of different components of the proton with increasing center-of-mass energy.
	
	\subsection{Comparison with Other Models}
	At the extrapolated energies of ${\sqrt{s}}$ = 14, 15, and 28 TeV, which is beyond the capabilities of LHC, the results of fitting our models show good agreement with the differential cross sections fits calculated by the extended Bialas-Bzdak (BB) model of ref. \cite{Nemes:2015iia}. The BB model treats the proton as a composite object with a real part in the scattering amplitude satisfying unitarity, and offers predictions that are sensitive to the spatial configuration within the proton. Our models successfully fitted both the diffractive cone and the first diffractive minimum regions, mirroring the non-trivial, non-exponential features in the differential cross-section as predicted by their approach. In our models, the presence of multiple exponential terms with varying weights and slopes is responsible for producing the interference patterns and diffractive features related to the predictions of the BB model formalism at the extrapolated LHC energies. At the 27.43 GeV data, our models showed good fitting agreement with the fitting result of the Froissaron and Maximal Odderon (FMO) model of ref. \cite{Bence:2020usl}, which stems from analyticity, crossing symmetry, and unitarity constraints, introduces asymptotically growing even-under-crossing and odd-under-crossing components in the amplitude. This model employed simplifying assumptions to study the role of Froissaron and Maximal Odderon in spin phenomena during elastic scattering. The FMO model holds the Froissaron responsible for the rise in the total cross section, and the Maximal Odderon contribution occurs with a distinct behavior in the real part of the amplitude. Though our models may not explicitly separate odd and even components, the successful fit of our models to the data where Odderon effects are most visible implies that the superposition of exponential forms can accommodate such subtleties indirectly through effective combinations of parameters. At 200 GeV and 800 GeV, our models showed a strong agreement with the predicted differential cross section fit results by the improved impact picture framework of ref. \cite{Bourrely:1978da}. This framework incorporates a refined matter distribution tied to the charge distribution of the proton, Regge background contributions, and a non-trivial hadronic matter current. The impact picture model refines the matter distribution by tying it to the known electromagnetic charge distribution of the proton, accounts for the spatial overlap and opacity function in the transverse impact parameter space. The use of multiple exponentials in our models allows for an approximate inverse Fourier representation of complex impact parameter distributions. The agreement of proposed models with the differential cross-section data (particularly the forward peak, dip structure, and low-${\mid t\mid}$ behavior) confirms that these models can be used as phenomenological tools to understand the dynamics of the proton.
	
	\subsection{Total Elastic Cross Section}
	The total elastic cross section ${\sigma_{\text{el}}}$ is calculated by numerically integrating the differential cross section models with the fitted parameters over all energies by the following equation.
	\begin{equation}
		\sigma_{\text{el}} = \int_{t_{\text{min}}}^{t_{\text{max}}} \frac{d\sigma}{dt} dt
	\end{equation}
	For all the TeV and GeV energies the $\frac{d\sigma}{dt}$ models are integrated in ${\mid t\mid}$-range of ${0 \leq {\mid t\mid} \leq 4.2}$ ${\textrm{GeV}^{2}}$ and ${0 \leq {\mid t\mid} \leq 12}$ ${\textrm{GeV}^{2}}$ respectively. The resulting values of ${\sigma_{\text{el}}}$, which are shown in Table IX, show good consistency with experimental measurements where available.
	
	\renewcommand{\arraystretch}{1.5}
	\setlength{\tabcolsep}{8pt}
	\begin{table}[h]
		\centering
		\caption{Results of elastic cross section ${\sigma_{\text{el}}}$ by integration of our models of the differential cross section.}
		\label{tab:parameters}
		\scalebox{0.60}{
			\begin{tabular}{c|cccccccccccc}
				\toprule
				\textbf{${\boldsymbol{\sqrt{s}}}$} & Model 1 & Model 2 & Model 3 & Model 4 & Model 5 & Model 6 & Model 7 & ISR \cite{Baksay:1978sg} & 2.76 TeV \cite{Cafagna:2021sge} & 7 \textrm{TeV} \cite{TOTEM:2013lle} & 8 \textrm{TeV} \cite{PhysRevLett.111.012001} & 13 \textrm{TeV} \cite{GAnchtev2019} \\
				\midrule
				& ($mb$) & ($mb$) & ($mb$) & ($mb$) & ($mb$) & ($mb$) & ($mb$) & ($mb$) & ($mb$) & ($mb$) & ($mb$) & ($mb$) \\
				\midrule
				23 ${\textrm{GeV}}$ & 6.5266 & 6.32625 & 6.60557 & 6.71209 & 6.1923 & 6.39853 & 6.39189 & - & - & - & - & - \\
				23.5 ${\textrm{GeV}}$ & 6.63969 & 6.66799 & 6.64812 & 6.66571 & 6.51434 & 6.53946 & 6.63523 & 6.82 & - & - & - & - \\
				27.43 ${\textrm{GeV}}$ & 7.14003 & 7.71544 & 7.34718 & 7.05101 & 7.61527 & 7.54103 & 7.0634 & - & - & - & - & - \\
				30.7 ${\textrm{GeV}}$ & 7.38816 & 7.15672 & 7.07817 & 7.00381 & 7.66528 & 7.86892 & 7.29245 & 7.39 & - & - & - & - \\
				44.7 ${\textrm{GeV}}$ & 7.21617 & 7.6702 & 7.34404 & 7.07463 & 7.61901 & 7.70186 & 7.44081 & 7.45 & - & - & - & - \\
				52.8 ${\textrm{GeV}}$ & 7.23764 & 7.40069 & 7.27359 & 7.35688 & 7.20325 & 7.37242 & 7.47945 & 7.56 & - & - & - & - \\
				62.5 ${\textrm{GeV}}$ & 7.54864 & 7.72488 & 7.7626 & 7.71417 & 7.0159 & 7.09113 & 7.85378 & 7.77 & - & - & - & - \\
				200 ${\textrm{GeV}}$ & 10.11363 & 10.0696 & 10.0022 & 10.1906 & 10.2148 & 10.247 & 10.3973 & - & - & - & - & - \\
				800 ${\textrm{GeV}}$ & 12.08162 & 12.0289 & 12.067 & 12.0671 & 12.0981 & 12.0553 & 12.0042 & - & - & - & - & - \\
				2.76 ${\textrm{TeV}}$ & 21.7784 & 21.4749 & 20.7911 & 22.7031 & 21.6311 & 21.5425 & 23.2111 & - & 21.8 ± 1.4 & - & - & - \\
				7 ${\textrm{TeV}}$ & 23.4114 & 25.4306 & 24.4253 & 24.3276 & 24.6489 & 24.3124 & 27.6421 & - & - & 25.4 ± 1.1 & - & - \\
				8 ${\textrm{TeV}}$ & 26.6814 & 26.522 & 27.4732 & 26.554 & 26.8882 & 27.6492 & 27.1972 & - & - & - & 27.1 ± 1.4 & - \\
				13 ${\textrm{TeV}}$ & 31.0991 & 31.6471 & 31.5338 & 31.4935 & 31.459 & 31.7374 & 30.9749 & - & - & - & - & 31.0 ± 1.7 \\
				14 ${\textrm{TeV}}$ & 32.7335 & 32.0887 & 32.3486 & 32.44 & 31.2397 & 32.0094 & 32.5215 & - & - & - & - & - \\
				15 ${\textrm{TeV}}$ & 34.1479 & 34.1934 & 34.3307 & 34.7952 & 34.2409 & 34.3047 & 34.607 & - & - & - & - & - \\
				28 ${\textrm{TeV}}$ & 35.6335 & 36.0963 & 36.3509 & 36.0248 & 37.1658 & 36.2268 & 36.9525 & - & - & - & - & - \\
				\bottomrule
			\end{tabular}
		}
	\end{table}
	Notably, ${\sigma_{\text{el}}}$ increases with energy, reflecting the growing blackness of the proton. At ISR energies (23 GeV - 62.5 GeV), ${\sigma_{\text{el}}}$ remains relatively small ($\sim$7–10 mb). At LHC energies (2.76 TeV - 13 TeV), ${\sigma_{\text{el}}}$ rises significantly ($\sim$25–30 mb). Extrapolated values at 14, 15, and 28 TeV predict further moderate increases, which are consistent with such an increasing trends reported by the TOTEM and COMPETE Collaborations \cite{COMPETE2002}\cite{Jenkovszkye24071001}. Accurate determination of ${\sigma_{\text{el}}}$ is critical for understanding various quantities and physical features at ultra-high energies.\\
	\renewcommand{\arraystretch}{1.5}
	\setlength{\tabcolsep}{8pt}
	\begin{table}[h]
		\centering
		\caption{Relative error calculation of the $\sigma_{el}$ results}
		\label{tab:parameters}
		\scalebox{0.600}{
			\begin{tabular}{c|ccccccc}
				\toprule
				\textbf{${\boldsymbol{\sqrt{s}}}$} & Model 1 & Model 2 & Model 3 & Model 4 & Model 5 & Model 6 & Model 7 \\
				\midrule
				 & \%\space Abs. Rel. Error & \%\space Abs. Rel. Error  & \%\space Abs. Rel. Error & \%\space Abs. Rel. Error & \%\space Abs. Rel. Error & \%\space Abs. Rel. Error & \%\space Abs. Rel. Error\\
				\midrule
				23.5 ${\textrm{GeV}}$ & 2.6438 & 2.2289 & 5.1918 & 2.2623 & 4.4818 & 4.1135 & 2.7092 \\
				30.7 ${\textrm{GeV}}$ & 0.0249 & 3.1567 & 4.2196 & 5.2259 & 3.7250 & 6.4807 & 1.3200 \\
				44.7 ${\textrm{GeV}}$ & 3.1387 & 2.9557 & 1.4223 & 5.0385 & 2.2686 & 3.3807 & 0.1234 \\
				52.8 ${\textrm{GeV}}$ & 4.2640 & 2.1073 & 3.7885 & 2.6868 & 4.7189 & 2.4812 & 1.0655 \\
				62.5 ${\textrm{GeV}}$ & 2.8489 & 0.5807 & 0.0952 & 0.7185 & 9.7053 & 8.7371 & 1.0783 \\
				2.76 ${\textrm{\textrm{TeV}}}$ & 0.0991 & 1.4913 & 4.6280 & 4.1427 & 0.7748 & 1.1812 & 6.4729 \\
				7 ${\textrm{\textrm{TeV}}}$ & 7.8291 & 0.1205 & 3.8374 & 4.2221 & 2.9571 & 4.2819 & 8.8272 \\
				8 ${\textrm{\textrm{TeV}}}$ & 1.5446 & 2.1328 & 1.3771 & 2.0148 & 0.7817 & 2.0266 & 0.3587 \\
				13 ${\textrm{\textrm{TeV}}}$ & 0.3197 & 2.0874 & 1.7219 & 1.5919 & 1.4807 & 2.3787 & 0.0811 \\
				\bottomrule
			\end{tabular}
		}
	\end{table}

An absolute relative error (ARE) estimation of $\sigma_{el}$ is performed for each model across the energy range (23.5 GeV - 13 TeV). The calculated ARE values are mentioned in Table X. For the relative error, we used the following equation.
\begin{equation}
	\partial\sigma = \frac{\mid \sigma_{model}-\sigma_{exp} \mid}{\sigma_{exp}} \times 100
\end{equation}
Where $\partial\sigma$ is the relative percentage absolute error or percentage deviation from the reference value. $\sigma_{model}$ and $\sigma_{exp}$ are the model and experimental values of the elastic cross section. The model 1 results of $\sigma_{el}$ showed the relative error in the range of ${0.0249\%-7.8291\%}$. The model 2 results showed the error in the range of ${0.1205\%-3.1567\%}$. The model 3 results showed the error in the range of ${0.0952\%-5.1918\%}$. The model 4 results showed the error in the range ${0.7185\%-5.2259\%}$. The model 5 results showed the error in the range of ${0.7748\%-9.7053\%}$. The model 6 results gave the error in the range of ${1.1812\%-8.7371\%}$. And the model 7 showed the relative error in the range of ${0.0811\%-8.8272\%}$. The good fit range of the relative error (ARE) for high precision models is usually considered to be ${ARE < 5\%}$. For global fits, especially when covering a broad energy range (from ISR to LHC), the ARE range is ${5\% \leq ARE \leq 10\%}$. Poor Fit Range corresponds to the ${ARE > 15\%}$, which indicates significant disagreement with data. In this study, the results of the total elastic cross section for all seven models are found to be in the ARE range of ${0.0249\% \leq ARE \leq 9.7053\%}$. The lowest relative error range is found across the model 2 results of the total elastic cross section $\sigma_{el}$.

\subsection{Total and Inelastic Cross Section}
We extend the calculations of this study for the observable, total cross section ${\sigma_{\text{tot}}}$, using the optical theorem \cite{TOTEM:2013lle}. The results of the total cross section ${\sigma_{\text{tot}}}$ are obtained by the extrapolation of the models, at $t=0$:
\begin{equation}
	\sigma_{\text{tot}}^{2} = \frac{16 \pi (\hbar c)^{2}}{1 + \rho^{2}} (\frac{d\sigma}{dt})_{t=0}
\end{equation}
\begin{equation}
	\space \sigma_{\text{inel}} = \sigma_{\text{tot}} - \sigma_{\text{el}}
\end{equation}
The conversion factor $(\hbar c)^{2}\simeq 0.38 \textrm{GeV}^{2} mbarn$ is used in these ${\sigma_{\text{tot}}}$ calculations. The $\rho$ parameter values used for each energy, along with the calculated total cross sections, are given in Table XI. The $\rho$ parameter values used in this work are taken from refs. \cite{Amaldi1980,star2020,Cafagna:2021sge,TOTEM:2013lle,PhysRevLett.111.012001,GAnchtev2019,Nemes:2015iia} for all energies except for 27.43 GeV and 800 GeV. For 27.43 GeV, $\rho$ is assumed equal to 0.033, the average of $\rho$ values for 23.5 GeV and 30.7 GeV. $\rho$ parameter value for 800 GeV is assumed to be little bit greater than 0.12 ($\rho$ parameter value for 200 GeV). The calculations of the ${\sigma_{\text{tot}}}$ relied mostly on the outcome of the models after their extrapolation at $t=0$ with the determined fitted parameter values. The models 5 and 6 contained the power law factor in the first exponential term, due to which the extrapolated values of the $(d\sigma /dt)_{t=0}$ are drastically lowered and result in very low values of total cross section ${\sigma_{\text{tot}}}$. When this factor is neglected in these models, then ${\sigma_{\text{tot}}}$ values increase while keeping other parameters the same. The $\sigma_{tot}$ results of models 5 and 6 have been shown after doing this modification in Table XI. The calculated results of the inelastic cross section ${\sigma_{\text{inel}}}$ are given in Table XI. Comparison plots of ${\sigma_{\text{el}}}$, ${\sigma_{\text{tot}}}$, and ${\sigma_{\text{inel}}}$ are shown in Figure 8 for graphical visualization purposes of the results of these observables. The ${\sigma_{\text{el}}}$, ${\sigma_{\text{tot}}}$ reference curves are obtained by joining the experimental reference values (23.5 GeV, 30.7 GeV, 44.7 GeV, 52.8 GeV, 62.5 GeV, 2.76 TeV, 7 TeV, 8 TeV, and 13 TeV energy domains) of the Tables IX and XI. The difference of these reference values is used to obtain a reference curve for the inelastic cross section. It is evident from Figure 8 that ${\sigma_{\text{el}}}$ results from our models show good agreement with ${\sigma_{\text{el}}}$ reference values, and show quite reasonable values at the extrapolated energies. All the calculated ${\sigma_{\text{el}}}$ values lie on the reference curve, depicting very small or almost no major difference with the reference points. Generally, ${\sigma_{\text{tot}}}$ results of the models 1, 2, 3, 4, and 7 showed a lesser difference with the experimentally measured reference values as compared to the models 5 and 6, which have shown the least agreement among all the models. The results across models 5 and 6 show the highest underestimation of ${\sigma_{\text{tot}}}$ and are observed at the lowest level below the reference curve. The models 1, 2, 3, and 4 underestimated the total cross sections from c.m. energies 23 GeV - 62.5 GeV. And from c.m. energies 200 GeV to 28 TeV, these models showed more agreement with the reference values. Among the models 1, 2, 3, and 4, model 3 results showed the least agreement at c.m energies 200 GeV to 28 TeV with the reference values. The results of models 1, 4, and 7 showed the least difference from the reference values among all the models, with model 7 showing the most agreement with reference values of ${\sigma_{\text{tot}}}$, and its values almost appear on the reference curve. For the observable ${\sigma_{\text{inel}}}$, a similar trend can be observed in Figure 8 across all the model results with the reference values. Among these models, the model 7 results for ${\sigma_{\text{inel}}}$ are in most agreement with the reference values. It is important to observe that the calculated results showed a rise in the observables, ${\sigma_{\text{el}}}$, ${\sigma_{\text{tot}}}$, and ${\sigma_{\text{inel}}}$ with increasing c.m. energy for all the models. It is evident from Figure 8 that ${\sigma_{\text{el}}}$ rises slowly as compared to the ${\sigma_{\text{tot}}}$ with increasing c.m. energies. Subject to model-dependent theoretical interpretations, the ratio ${\sigma_{\text{el}}}/{\sigma_{\text{tot}}}$ can provide some information on the shape and opacity of the proton. The consistent increase in this ratio with energy is frequently considered to be an indication of an increase in proton size and opacity with energy \cite{Adamczyk:2015gfy}. The calculations of ratio, ${\sigma_{\text{el}}}/{\sigma_{\text{tot}}}$ by TOTEM have been reported in ref. \cite{TOTEM:2013vij}. Results of ${\sigma_{{el}}}$ and ${\sigma_{{tot}}}$ by our models are analyzed in this study for ${\sigma_{\text{el}}}/{\sigma_{\text{tot}}}\sim (log(s))^{-n}$ behavior which is given in ref. \cite{Collins:1977jy} for elastic processes, with n=1, and is related to the shrinkage of the forward peak, described by approximating the residue by a single exponential to obtain the differential cross section that has a similar s dependence which is introduced in our models. This ratio suggests that the elastic fraction of the total scattering decreases with increasing c.m. energy, indicating the decrease of purely elastic events, which never entirely vanish, giving an indication that inelastic processes dominate very high energies. At GeV energies (23 GeV - 800 GeV), the elastic cross-section comprises ${\sim 15–20 \%\space}$ of the total cross-section. The ratio fits well with the ${\sim log(s)}$ behavior. At LHC energies 2.76 TeV - 13 TeV, the ratio lies in the range ${{\sigma_{\text{el}}}/{\sigma_{\text{tot}}} \sim 0.23-0.27}$. Our results for the behavior, ${\sigma_{\text{el}}}/{\sigma_{\text{tot}}}\sim (log(s))^{-1}$ show consistency especially for the model 7 at LHC energies of 2.76 TeV - 13 TeV and at the extrapolated energies of 14 TeV, 15 TeV and 28 TeV. This energy regime is very high so that the scattering is dominated by the exchange of Pomeron that contributes to a logarithmic rise in ${\sigma_{\text{tot}}}$, which is in accordance with the Froissart-Martin bound. However, at GeV energies, which are much lower than LHC energies, the ratio ${\sigma_{\text{el}}}/{\sigma_{\text{tot}}}$ shows inconsistency with the $(log(s))^{-1}$ behavior.
In this study, the deviations in ${\sigma_{\text{tot}}}$ and ${\sigma_{\text{inel}}}$ values from their reference values, which have been observed in the results of the models can be attributed to the lack of explicit constraints in the forward scattering region in the models. It is usually found that model-independent parametrizations, which are made to fit the mid-$\mid t\mid$ region, may not constrain the forward limit effectively, which can affect the calculations of ${\sigma_{\text{tot}}}$ and ${\sigma_{\text{inel}}}$. Also, the model parameter values have been obtained through non-linear fitting of the models with the data, with no initial values, which leads to many solutions, subsequently affecting the results of the observables found by the models. We adopted the simple parameter choices for the functions A(s,t) and B(s,t) for t dependence in our models to reduce the number of free parameters and for other reasons, which resulted in oversimplification in terms of t dependence, which affected the extrapolation at t = 0. Moreover, it can be understood that modifications related to certain physical constraints are required in the composite exponential structures of the models to better represent the scattering amplitude contributions, thereby overcoming the limitations in the calculations of $\sigma_{tot}$ and $\sigma_{el}$ observables. These are also required for correlation with other models and other advances in the modeling of the differential cross section. In the present study, we are not interested in investigating the energy dependence of other free parameters of our models. It can be very significant to investigate the scaling patterns in the free parameter values obtained in this study. Moreover, it will be very useful to study how their extracted values relate to energy and other physical parameters used in other theoretical and phenomenological approaches.

\begin{figure}[h]
	\centering
	\includegraphics[width=0.85\textwidth]{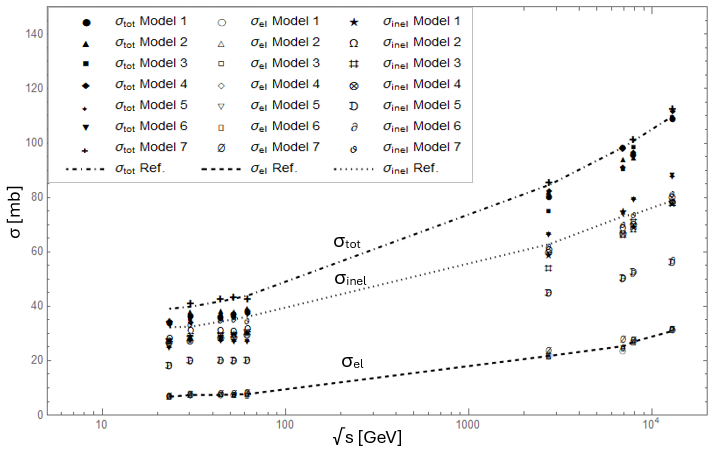}
	\caption{Graphical comparison of total, elastic, and inelastic cross-section calculation results by our models presented with the reference values (as shown in Tables IX, XI, and XII) as a function of the centre-of-mass energy. The results across 23.5 GeV, 30.7 GeV, 44.7 GeV, 52.8 GeV, 62.5 GeV, 2.76 TeV, 7 TeV, 8 TeV, and 13 TeV energy domains are shown. The dotted dashed line represents the reference curve of $\sigma_{\text{tot}}$. The dotted line represents the reference curve of $\sigma_{\text{inel}}$. And the dashed line represents the reference curve of $\sigma_{\text{el}}$. These reference curves are obtained by joining only the reference points at the aforementioned energies.}
	\label{fig:cross-section}
\end{figure}

	\renewcommand{\arraystretch}{1.5}
	\setlength{\tabcolsep}{8pt}
	\begin{table}[h]
		\centering
		\caption{Results of the total cross section ${\sigma_{\text{tot}}}$ by extrapolation of our models at $t=0$.}
		\label{tab:parameters}
		\scalebox{0.57}{
			\begin{tabular}{c|c|ccccccccccccc}
				\toprule
				\textbf{${\boldsymbol{\sqrt{s}}}$} & $\rho$ & Model 1 & Model 2 & Model 3 & Model 4 & Model 5 & Model 6 & Model 7 & ISR \cite{Baksay:1978sg} & 2.76 \textrm{TeV} \cite{Cafagna:2021sge} & 7 \textrm{TeV} \cite{TOTEM:2013lle} & 8 \textrm{TeV} \cite{PhysRevLett.111.012001} & 13 \textrm{TeV} \cite{GAnchtev2019} & \cite{Nemes:2015iia} \\
				\midrule
				&  & $\sigma_{tot}$ & $\sigma_{tot}$ & $\sigma_{tot}$ & $\sigma_{tot}$ & $\sigma_{tot}$ & $\sigma_{tot}$ & $\sigma_{tot}$ & $\sigma_{tot}$ & $\sigma_{tot}$ & $\sigma_{tot}$ & $\sigma_{tot}$ & $\sigma_{tot}$ & $\sigma_{tot}$\\
				\midrule
				&  & ($mb$) & ($mb$) & ($mb$) & ($mb$) & ($mb$) & ($mb$) &  ($mb$) & ($mb$) & ($mb$) & ($mb$) & ($mb$) & ($mb$) & ($mb$) \\
				\midrule
				23 ${\textrm{GeV}}$ & 0.020 \cite{Amaldi1980} & 33.8850 & 28.027 & 33.7003 & 33.4029 & 23.9496 & 24.6002 & 35.1545 & - & - & - & - & - & - \\
				23.5 ${\textrm{GeV}}$ & 0.025 \cite{Amaldi1980} & 34.6341 & 35.3842 & 34.4375 & 34.1314 & 25.1989 & 25.2510 & 34.6533 & 39.13 $\pm$ 0.32 & - & - & - & - & - \\
				27.43 ${\textrm{GeV}}$ & 0.033 & 36.6113 & 38.2435 & 36.4461 & 34.9026 & 32.5592 & 27.2584 & 37.7829 & - & - & - & - & - & - \\
				30.7 ${\textrm{GeV}}$ & 0.042 \cite{Amaldi1980} & 37.1976 & 38.8645 & 35.8101 & 34.9769 & 28.1613 & 28.0592 & 41.8842 & 39.91 $\pm$ 0.33 & - & - & - & - & - \\
				44.7 ${\textrm{GeV}}$ & 0.062 \cite{Amaldi1980} & 36.8438 & 39.0422 & 37.0792 & 35.8318 & 27.896 & 27.9117 & 43.5078 & 41.89 $\pm$ 0.32 & - & - & - & - & - \\
				52.8 ${\textrm{GeV}}$  & 0.078 \cite{Amaldi1980} & 37.2086 & 38.4814 & 37.0274 & 36.3357 & 27.3668 & 27.3472 & 43.3462 & 42.85 $\pm$ 0.33 & - & - & - & - & - \\
				62.5 ${\textrm{GeV}}$ & 0.095 \cite{Amaldi1980} & 37.9764 & 39.437 & 38.3695 & 37.5656 & 27.1073 & 27.1158 & 42.6843 & 44.0 & - & - & - & - & - \\
				200 ${\textrm{GeV}}$ & 0.12 \cite{star2020} & 50.1871 & 51.2587 & 49.7556 & 48.126 & 35.6001 & 35.4264 & 51.9174 & - & - & - & - & - & - \\
				800 ${\textrm{GeV}}$ & 0.122 & 61.4287 & 57.2562 & 61.2519 & 61.7630 & 40.5889 & 40.1546 & 61.7197 & - & - & - & - & - & - \\
				2.76 ${\textrm{TeV}}$ & 0.12 \cite{Cafagna:2021sge} & 80.1842 & 81.8242 & 74.4543 & 82.4611 & 66.2536 & 66.1486 & 85.1398 & - & 84.7 $\pm$ 3.3 & - & - & - & - \\
				7 ${\textrm{TeV}}$ & 0.14 \cite{TOTEM:2013lle} & 97.7588 & 93.8851 & 90.2326 & 91.0688 & 74.6628 & 74.1022 & 97.4241 & - & - & 98.6 $\pm$ 2.2 & - & - & - \\
				8 ${\textrm{TeV}}$ & 0.14 \cite{PhysRevLett.111.012001} & 95.1507 & 94.7947 & 97.8477 & 96.3823 & 78.8789 & 79.011 & 100.67 & - & - & - & 101.7 $\pm$ 2.9 & - & - \\
				13 ${\textrm{TeV}}$ & 0.098 \cite{GAnchtev2019} & 108.629 & 109.652 & 109.145 & 111.821 & 87.3737 & 88.138 & 112.366 & - & - & - & - & 110.6 $\pm$ 3.4 & - \\
				14 ${\textrm{TeV}}$ & 0.108 \cite{Nemes:2015iia} & 122.689 & 119.65 & 90.82 & 120.246 & 97.3395 & 94.9042 & 122.839 & - & - & - & - & - & 108.1 \\
				15 ${\textrm{TeV}}$ & 0.109 \cite{Nemes:2015iia} & 123.460 & 121.929 & 92.588 & 123.991 & 100.533 & 102.14 & 124.913 & - & - & - & - & -& 109.1 \\
				28 ${\textrm{TeV}}$ & 0.114 \cite{Nemes:2015iia} & 134.549 & 132.356 & 102.644 & 134.465 & 108.042 & 110.617 & 132.873 & - & - & - & - & - & 118.5 \\
				\bottomrule
			\end{tabular}
		}
	\end{table}
		\renewcommand{\arraystretch}{1.5}
	\setlength{\tabcolsep}{8pt}
	\begin{table}[h]
		\centering
		\caption{Results of inelastic cross section ${\sigma_{\text{inel}}}$.}
		\label{tab:parameters}
		\scalebox{0.60}{
			\begin{tabular}{c|cccccccccccc}
				\toprule
				\textbf{${\boldsymbol{\sqrt{s}}}$} & Model 1 & Model 2 & Model 3 & Model 4 & Model 5 & Model 6 & Model 7 & ISR \cite{Baksay:1978sg} & 2.76 \textrm{TeV} \cite{Cafagna:2021sge} & 7 \textrm{TeV} \cite{TOTEM:2013lle} & 8 \textrm{TeV} \cite{PhysRevLett.111.012001} & 13 \textrm{TeV} \cite{GAnchtev2019} \\
				\midrule
				& $\sigma_{inel}$ & $\sigma_{inel}$ & $\sigma_{inel}$ & $\sigma_{inel}$ & $\sigma_{inel}$ & $\sigma_{inel}$ & $\sigma_{inel}$ & $\sigma_{inel}$ & $\sigma_{inel}$ & $\sigma_{inel}$ & $\sigma_{inel}$ & $\sigma_{inel}$ \\
				\midrule
				& ($mb$) & ($mb$) & ($mb$) & ($mb$) & ($mb$) & ($mb$) & ($mb$) & ($mb$) & ($mb$) & ($mb$) & ($mb$) & ($mb$) \\
				\midrule
				23 ${\textrm{GeV}}$ & 27.3584 & 21.7008 & 27.0947 & 26.6909 & 17.7573 & 18.2017 & 28.7626 & - & - & - & - & - \\
				23.5 ${\textrm{GeV}}$ & 27.9944 & 28.7162 & 27.7894 & 27.4657 & 18.6845 & 18.7116 & 28.0181 & 29.07 $\pm$ 0.32 & - & - & - & - \\
				27.43 ${\textrm{GeV}}$ & 29.4713 & 30.5281 & 29.0989 & 27.8516 & 24.9439 & 19.7174 & 30.7195 & - & - & - & - & - \\
				30.7 ${\textrm{GeV}}$ & 29.8095 & 31.7078 & 28.7319 & 27.9731 & 20.496 & 20.1903 & 34.5917 & 32.52 $\pm$ 0.33 & - & - & - & - \\
				44.7 ${\textrm{GeV}}$ & 29.6276 & 31.372 & 29.7351 & 28.7572 & 20.277 & 20.2098 & 36.067 & 34.44 $\pm$ 0.32 & - & - & - & - \\
				52.8 ${\textrm{GeV}}$ & 29.9709 & 31.0808 & 29.7538 & 28.9788 & 20.1635 & 19.9748 & 35.8667 & 35.29 $\pm$ 0.33 & - & - & - & - \\
				62.5 ${\textrm{GeV}}$ & 30.4277 & 31.7121 & 30.6069 & 29.8514 & 20.0914 & 20.0247 & 34.8305 & 36.23 $\pm$  & - & - & - & - \\
				200 ${\textrm{GeV}}$ & 40.0735 & 41.1891 & 39.7534 & 37.9354 & 25.3853 & 25.1794 & 41.5201 & - & - & - & - &  \\
				800 ${\textrm{GeV}}$ & 49.3471 & 47.1866 & 49.223 & 49.6959 & 28.4908 & 28.0993 & 49.7155 & - & - & - & - &  \\
				2.76 ${\textrm{TeV}}$ & 58.4058 & 60.3493 & 53.6632 & 59.758 & 44.6225 & 44.6061 & 61.9287 & - &  62.8 $\pm$ 2.9 & - & - & - \\
				7 ${\textrm{TeV}}$ & 74.3474 & 68.4545 & 62.8073 & 66.7412 & 50.0139 & 49.7898 & 69.782 & - & - &  70.5 $\pm$ 2.9 & - & - \\
				8 ${\textrm{TeV}}$ & 68.9011 & 68.2727 & 70.3745 & 69.8283 & 51.9907 & 51.3618 & 73.4731 & - & - & - & 73.74 $\pm$ 0.09 & - \\
				13 ${\textrm{TeV}}$ & 77.5295 & 78.0051 & 77.6109 & 80.2876 & 55.9147 & 56.4006 & 81.3911 & - & - & - & - & 79.1 $\pm$ 1.8 \\
				14 ${\textrm{TeV}}$ & 89.9554 & 87.5616 & 58.4715 & 87.8062 & 66.0998 & 62.8948 & 90.3177 & - & - & - & - & - \\
				15 ${\textrm{TeV}}$ & 89.3124 & 87.5987 & 58.2573 & 89.1961 & 66.2917 & 67.8348 & 90.3059 & - & - & - & - & - \\
				28 ${\textrm{TeV}}$ & 98.9157 & 96.2595 & 66.2927 & 98.4403 & 70.8761 & 74.3899 & 95.9204 & - & - & - & - & - \\
				\bottomrule
			\end{tabular}
		}
	\end{table}
	The models of $\frac{d\sigma}{dt}$ used in this study can be used to predict several important observables that are very significant in understanding proton dynamics and the structure of the proton in $pp$ elastic scattering at high energies. For instance, the local slope parameter or logarithmic slope B(t) of the differential cross section, defined by taking s fixed at $s = s_0$ in our models, which are s and t-dependent. B(t) is obtained as
	\begin{equation}
		B(t) = \frac{d}{dt} \left[ \ln \left( |\frac{d\sigma (s,t)}{dt}|_{s=s_0} \right) \right],
	\end{equation}
	which reflects the \(t\)-dependent variation of the elastic scattering. This t-dependent slope parameter for $pp$ and $p\bar{p}$ data has been used in ref. \cite{TOTEM_2018psk} to investigate Odderon effect by finding $B(t)_{pp} \neq B(t)_{p\bar{p}}$. In ref. \cite{Csorgo:2021v}, an indication of two substructures of different sizes in proton is given at low and high energy domains by the t-dependent slope parameter. By the directions followed in ref. \cite{Csorgo:2021v}, the B(s) slope parameter can be taken as $B(s) \equiv B(s,t=0)$ which can be related to $\rho$ parameter, $\sigma_{el}$ and $\sigma_{{tot}}$ thereby leading to estimate scaling properties of shadow profile which are very significant to study structure and dynamics of proton at high energies. These B(s) and B(t) parameters obtained by our models can offer insights into changes in interaction geometry with their dependence on Mandelstam variables s and t, respectively. The increase in the slope parameter with ${\sqrt{s}}$ as a feature of $\frac{d\sigma}{dt}$ is linearly related to the radius of proton \cite{Antchev_2011}. The $\frac{d\sigma}{dt}$ models can also be used to predict the differential curvature and dip behavior by estimating the second derivative of \( \ln(d\sigma/dt) \) by keeping s fixed at $s=s_0$, thereby estimating C(t) which gives access to effects related to curvature of the cross section \cite{Csorgo2023}:
	\begin{equation}
		C(t) = \frac{d^2}{dt^2} \left[ \ln \left( |\frac{d\sigma(s,t)}{dt}|_{s=s_0} \right) \right],
	\end{equation}
	The C(t) parameter is very important in analysis of the dip–bump region and deviations from simple exponential behavior.
	
	\section{Conclusion}
	\label{sec:conclusion}
	Our primary motivation for investigating the inverse problem as a source of experimental information, suitable for model development, and potential linkages with QCD evolved from the absence of a pure QCD description of the $pp$ elastic scattering data and well-defined physical scenarios associated with a wide range of phenomenological models. For this purpose, we have performed a global comparative analysis of the elastic $pp$ scattering across a wide range of energies, extending from the lower ISR energies to the higher LHC energies and at the extrapolated energies, by fitting seven composite models of $\frac{d\sigma}{d|t|}$. Each model incorporated distinct exponential structures to represent the composite behavior of the differential cross section across different kinematic regions from the forward peak at small-${\mid t\mid}$, through the diffractive dip, to the high-${\mid t\mid}$ tail dominated by hard scattering processes. Our model fits reproduced the important features such as the forward exponential fall-off and the dip-bump structure. The behavior of the dip is well produced in the graphical representation of the data by the fits of our models. The dip position is found to be moving towards smaller ${\mid t\mid}$ values and becoming shallower with increasing center-of-mass energy. This physical trend indicates the increasing blackness and size of the proton at higher energies, consistent with the expected gluonic saturation effects and the shrinkage of the forward peak. The high-${\mid t\mid}$ behavior, which is governed by the inner quark structure, is also accurately described in the multi-exponential contributions of our models. The calculated total elastic cross sections, evaluated through numerical integration of the fitted models, showed good consistency with known measurements and with established predictions for the ISR, TOTEM experimental data, and also at extrapolated energies. The results of the total ${\sigma_{\text{tot}}}$ and inelastic cross sections ${\sigma_{\text{inel}}}$ are found using the proposed models.
It is observed that our models give results comparable to experimental data.
The rise in elastic, inelastic, and total cross sections is observed in their estimations by our models with increasing center of mass energies in both GeV and TeV energy domains. The analysis of the ratio ${\sigma_{\text{el}}}/{\sigma_{\text{tot}}}$ showed that our results produced expected patterns at GeV and TeV energy domains. However, the calculations of  ${\sigma_{\text{tot}}}$ and ${\sigma_{\text{inel}}}$ showed differences with the calculations by other models, which indicated the limitation of some of our models. However, it is crucial to consider these limitations as suggestions for developments and improvements. Based on the considerations related to the extrapolations of these models, these models can be considered in three groups. Group 1 contains models 1-4, group 2 contains models 5 and 6, and group 3 contains model 7. It is found that $\chi^{2}$ values (except at 200 GeV and 800 GeV) are very small which shows the agreement of proposed models with data. The similar agreement is observed between models and date in the graphical representation. The calculated ${\sigma_{\text{el}}}$ values are found very close to the experimental measured reference values. Each group has a different level of compatibility with the reference values of ${\sigma_{\text{tot}}}$ and ${\sigma_{\text{inel}}}$. On basis of the fitting agreement with all the data, lesser difference with reference values (explained in section \uppercase{iii}) of ${\sigma_{\text{el}}}$, ${\sigma_{\text{tot}}}$ and ${\sigma_{\text{inel}}}$ at most of the energies, and other considerations discussed in section \uppercase{iii}, the group 3 (model 7) is found to be the most reasonable among all the models. It is understood from the present analysis that an efficient modeling strategy demands detailed investigation of different analyses and results. Generally, the good fitting agreement of our models with previous significant works that include those based on the Bialas Bazdak (BB) model, impact picture, and the FMO approach, emphasizes the physical relevance of our parametrizations of the differential cross section. It reinforces the understanding that the internal structure of the proton and the dynamics of elastic scattering can be consistently described by modeling the differential cross sections. This study, therefore, provides a basis for future refinements of phenomenological models and offers predictive strength for upcoming experiments at higher and intermediate energies.
	
\section{Recommendations and Future Work}
	In the present study, the composite exponential models, which have been used to effectively describe the differential cross section behavior of $pp$ elastic scattering across a wide energy range, can be further improved in the directions of Section \uppercase{iii}. The parameters can be modified in terms of Mandelstam variables s and t, especially by developing parametrizations that have more effective theoretical roots. It is recommended to combine phenomenological fitting of these models with theoretical models such as eikonal frameworks and QCD-based saturation approaches to explain the underlying dynamics. These models can be considered for further validation with the availability of new data from the LHC and future colliders like the FCC, particularly at very low and very high-${\mid t\mid}$. The use of machine learning techniques may also offer improved predictive capabilities. We are hopeful that extensions of this study along the highlighted directions can promise richer insights into hadronic physics.


\section*{References}
	
	\begin{thebibliography}{99}
	
	\bibitem{DONNACHIE1992227}
	A. Donnachie and P.V. Landshoff, (1992) "Total cross sections", Phy. Lett. B \textbf{296}, 227-232.
	
	\bibitem{PhysRevD.83.077901}
	Block, M. M. and Halzen, F., (2011) "Forward hadronic scattering at 7 TeV: An update on predictions for the LHC", Phys. Rev. D \textbf{83}.
	
	\bibitem{DONNACHIE2013500}
	A. Donnachie and P.V. Landshoff, (2013) "${pp}$ and ${\bar{p}p}$ total cross sections and elastic scattering", Phys. Lett. B \textbf{727}, 500-505.
	
	\bibitem{Donnachie:2002en}
	Donnachie, S. and Dosch, G. and Landshoff, P. and Nachtmann, O., (2002) "Pomeron Physics and QCD", Cambridge University Press.
	
	\bibitem{Forshaw:1997dc}
	Forshaw, J.R. and Ross, D.A., (1997) "Quantum Chromodynamics and the Pomeron", Cambridge University Press.
	
	\bibitem{NAGY1979221}
	E. Nagy et al., (1979) "Measurements of elastic proton-proton scattering at large momentum transfer at the CERN intersecting storage rings", Nucl. Phys. B \textbf{150}, 221-267.
	
	\bibitem{TOTEM_2018psk}
	Antchev, G. et al., (2020) "Elastic differential cross-section $d\sigma/dt$ at $\sqrt{s}$ = 2.76 TeV and implications on the existence of a colourless C-odd three-gluon compound state", Eur. Phys. J. C \textbf{80}, CERN-EP-2018-341, TOTEM-2018-002.
	
	\bibitem{Antchev_2013}
	Antchev, G. et al., (2013) "Measurement of proton-proton inelastic scattering cross-section at ${\sqrt{s}}$=7 TeV", Europhysics Letters \textbf{101}, 21003.
	
	\bibitem{Antchev_2011}
	The TOTEM Collaboration and Antchev, G. et al., (2011) "Proton-proton elastic scattering at the LHC energy of 7 TeV", Europhysics Letters \textbf{95}, 41001.
	
	\bibitem{TOTEM:2021imi}
	The TOTEM Collaboration and Antchev, G. et al., (2022) "Characterisation of the dip-bump structure observed in proton-proton elastic scattering at ${\sqrt{s}}$ = 8 TeV", Eur. Phys. J. C \textbf{82}, 263.
	
	\bibitem{Antchev_2019}
	Antchev, G. et al., (2019) "Elastic differential cross-section measurement at ${\sqrt{s}}$=13 TeV by TOTEM", Eur. Phys. J. C \textbf{79}, 861.
	
	\bibitem{PhysRevLett.127.062003}
	Abazov, V. M. et al., (2021) "Odderon Exchange from Elastic Scattering Differences between ${pp}$ and ${\bar{p}p}$ Data at 1.96 TeV and from ${pp}$ Forward Scattering Measurements", Phys. Rev. Lett. \textbf{127}, 062003.
	
	\bibitem{Bozzo1984LowMT}
	Marco Bozzo et al., (1984) "Low momentum transfer elastic scattering at the CERN proton-antiproton collider", Phys. Lett. B \textbf{147}, 385-391.
	
	\bibitem{UA41985}
	UA4 Collaboration, (1985) "Elastic Scattering at the CERN SPS Collider Up to a Four Momentum Transfer of 1.55 GeV$^2$", Phys. Lett. B \textbf{155}, 197.
	
	\bibitem{CDF1994}
	CDF Collaboration, (1994) "Measurement of small angle $\bar{p}p$ elastic scattering at $\sqrt{s} = 546$ GeV and 1800 GeV", Phys. Rev. D, \textbf{50}, 5518.
	
	\bibitem{E7101988}
	E710 Collaboration, (1988) "Measurement of b, the Nuclear Slope Parameter of the $\bar{p} p$ Elastic Scattering Distribution at $\sqrt{s} = 1800$ GeV", Phys. Rev. Lett., \textbf{61}, 525.
	
	\bibitem{E7101989}
	E710 Collaboration, (1989) "Measurement of the $\bar{p} p$ Total Cross-Section at $\sqrt{s} = 1.8$ TeV", Phys. Rev. Lett., \textbf{63}, 2784.
	
	\bibitem{E7101992}
	E710 Collaboration, (1992) "Measurement of $\rho$, the ratio of the real to imaginary part of the $\bar{p} p$ forward elastic scattering amplitude, at $\sqrt{s} = 1.8$ TeV", Phys. Rev. Lett., \textbf{68}, 2433.
	
	\bibitem{ATLAS2016}
	Aaboud, M. and et al., (2016) "Measurement of the total cross section from elastic scattering in pp collisions at $\sqrt{s} = 8$ TeV with the ATLAS detector", Phys. Rev. Lett., \textbf{117}, 182002.
	
	\bibitem{STAR2016}
	Adamczyk, L. and et al, (2016) "Single spin asymmetry $A_N$ in polarized proton–proton elastic scattering at $\sqrt{s} = 200$ GeV at RHIC", Phys. Rev. Lett., \textbf{719}, 62-69.
	
	\bibitem{Froissart1961}
	M. Froissart, (1961) "Asymptotic behavior and subtractions in the Mandelstam representation", Phys. Rev., \textbf{123}, 1053.
	
	\bibitem{Gauron1992}
	P. Gauron, L. Lukaszuk and B. Nicolescu, (1992) "Consistency of the maximal odderon approach with the QFT constraints", Phys. Lett. B, \textbf{294}, 298.
	
	\bibitem{Block1994}
	M. M. Block et al., (1994) "The High-energy behavior of the forward scattering parameters $\sigma_{\text{total}}$, $\rho$ and B", arXiv:hep-ph/9412306, 73-78.
	
	\bibitem{Cudell2002}
	J. R. Cudell et al, (2002) "Hadronic scattering amplitudes: Medium-energy constraints on asymptotic behavior", Phys. Rev. D, \textbf{65}, 074024.
	
	\bibitem{COMPETE2002}
	COMPETE Collaboration, (2002) "Benchmarks for the forward observables at RHIC, the Tevatron Run II and the LHC", Phys. Rev. D, \textbf{89}, 201801.
	
	\bibitem{TOTEM2018}
	TOTEM Collaboration, (2018) "First determination of the $\rho$ parameter at $\sqrt{s} = 13$ TeV — probing the existence of a colourless three-gluon bound state", arXiv:1812.04732 [hep-ex].
	
	\bibitem{MMIslam}
	M. M. Islam, (2018) "{https://slac.stanford.edu/econf/C111215/papers/islam.pdf}".
	
	\bibitem{DREMIN2013241}
	I.M. Dremin and V.A. Nechitailo, (2013) "Proton periphery activated by multiparticle dynamics", Nuc. Phys. A, \textbf{916}, 241-248.
	
	\bibitem{PhysRevD.108.034028}
	Nekrasov, M. L., (2023) "$pp$ elastic scattering at ISR and LHC energies", Phys. Rev. D, \textbf{108}, 034028.

	\bibitem{Gelis2010}
	Gelis, F. et al., (2010) "The Color Glass Condensate", Ann. Rev. Nucl. Part. Sci., \textbf{60}, 463-489.
	
	\bibitem{Saleem:1980hu}
	Saleem, M. and Fazal-e-Aleem, (1980),"Dipole Pomeron and Proton-Proton Elastic Scattering at High Energies", Austr. Jour. Phys., \textbf{33}, 481.
	
	\bibitem{Cahn:1980}
	R. N. Cahn, (1982),"Coulomb-Hadronic Interference in an Eikonal Model", Z. Phys. C., \textbf{15}, 253-260.
	
	\bibitem{Fagundes22013}
	D. A. Fagundes, M. J. Menon, (2012),"Total Hadronic Cross Section and Elastic Slope: An Almost Model-Independent Connection", Nucl. Phys. A, \textbf{880}, 1-15.
	
	\bibitem{Kundrat:1994}
	V. Kundrát and M. Lokajíček, (1992),"High-energy elastic hadron scattering in Coulomb and hadronic regions", Phys. Rev. D, \textbf{46}, 4087–4090.
	
	\bibitem{TOT2015527}
	Antchev G. et al., (2015),"Evidence for non-exponential elastic proton–proton differential cross-section at low ${\mid t\mid}$ and ${\sqrt{s}}$ = 8 TeV by TOTEM", Nucl. Phys. B, \textbf{899}, 527-546.
	
	\bibitem{TOTEM:2019}
	Antchev, G. and others (TOTEM Collaboration), (2019),"Elastic Differential Cross-Section Measurement at 13 TeV by TOTEM", Eur. Phys. J. C, \textbf{79}, 103.
	
	\bibitem{Kohara:2017aix}
	Kohara, A. K. and Ferreira, E. and Kodama, T., (2017), "Amplitudes for High-Energy Proton-Proton Elastic Scattering", Eur. Phys. J. C, \textbf{77}, 877.
	
	\bibitem{Jenkovszky:2018pcm}
	Jenkovszky, L'aszl'o and Szanyi, Istv'an, (2018), "Elastic and inelastic diffraction at the LHC", EPJ Web Conf., \textbf{172}, 06004.
	
	\bibitem{MARTYNOV2018414}
	E. Martynov, (2018), "Did TOTEM experiment discover the Odderon?", Phys. Lett. B, \textbf{778}, 414-418.
	
	\bibitem{HEGS1}
	O.V. Selyugin, (2015), "Nucleon structure and the high energy interactions", Phys. Rev. D, \textbf{91}, 113003.
	
	\bibitem{HEGS2}
	O.V. Selyugin, (2012), "GPDs of the nucleons and elastic scattering at high energies", Eur. Phys. J. C, \textbf{72}, 2073.
	
	\bibitem{HEGS4}
	O.V. Selyugin, (2019), "New feature in the differential cross sections at 13 TeV measured at the LHC", Phys. Lett. B, \textbf{797}, 134870.
	
	\bibitem{HEGS5}
	O.V. Selyugin, (2024), "New properties of elastic pp and $p\bar{p}$ scattering at high energies", Eur. Phys. J. C, \textbf{84}, 649.
	
	\bibitem{HoloQCD1}
	Xie. Wei,Watanabe. Akira, Huang. Mei, (2019), "Elastic proton-proton scattering at LHC energies in holographic QCD", J. of High Energy Phys., \textbf{2019}, 53.
	
	\bibitem{PHILLIPS1973412}
	R.J.N. Phillips and V. Barger, (1973), "Model independent analysis of the structure in ${pp}$ scattering", Phys. Lett. B, \textbf{46}, 412-414.
	
	\bibitem{Gonçalves1973}
	Gonçalves, V.P., Silva, P.V.R.G., (2015), "The Phillips–Barger model for the elastic cross section and the Odderon", Eur. Phys. J. C, \textbf{79}, 237.
	
	\bibitem{Nemes:2015iia}
	Nemes, F. and C. org, T. and Csanad, M., (2015), "Excitation function of elastic ${pp}$ scattering from a unitarily extended Bialas-Bzdak model", Int. J. Mod. Phys. A, \textbf{30}, 1550076.
	
	\bibitem{Csorgo2023}
	Csörgő, Tamás and Hegyi, Sandor and Szanyi, István, (2023), "Lévy $\alpha$-Stable Model for the Non-Exponential Low-$\mid t\mid$ Proton–Proton Differential Cross-Section", Universe, \textbf{9}, 361.
	
	\bibitem{Praszalowicz:2025djk}
	Praszalowicz, M. et al., (2025), "Scaling of the elastic proton-proton cross-section", arXiv.2501.08398[hep-ph].
	
	\bibitem{BALDENEGRO2024138960}
	Baldenegro, C. et al., (2024), "Scaling laws of elastic proton-proton scattering differential cross sections", Phys. Lett. B. \textbf{856}, 138960.
	
	\bibitem{Block2011}
	M. M. Block and F. Halzen, (2011), "New fit to high-energy $pp$ and $p\bar{p}$ scattering data", Phys. Rev. D, \textbf{83}, 077901.
	
	\bibitem{Anisovich2014}
	V. V. Anisovich et al., (2014), "Impact parameter analysis of pp and $p\bar{p}$ scattering", Phys. Rev. D, \textbf{90}, 074005.
	
	\bibitem{Dulat2016}
	S. Dulat et al., (2016), "New Parton Distribution Functions from the CT14 Global Analysis", Phys. Rev. D, \textbf{93}, 033006
	
	\bibitem{Antchev2013}
	TOTEM Collaboration, (2013), "Luminosity-independent measurement of the total proton-proton cross-section at $\sqrt{s} = 8$ TeV", Phys. Rev. Lett., \textbf{111}, 012001.
	
	\bibitem{Auger2015}
	Pierre Auger Collaboration, (2016), "Testing Hadronic Interactions at Ultra-High Energies", Phys. Rev. Lett., \textbf{117}, 192001.
	
	\bibitem{Ostapchenko2019}
	S. Ostapchenko, (2019), "QGSJET-II: physics, recent improvements, and results for air showers", EPJ Web Conf., \textbf{208}, 11002.
	
	\bibitem{Grau2018}
	Grau, A. et al., (2019), "The elastic differential pp cross-section at 13 TeV: an empirical model analysis", EPJ Web Conf., \textbf{206}, 06003.
	
	\bibitem{Dremin20131}
	Dremin, I. M., (2013), "Elastic pp scattering from the optical point to past the dip: An empirical model analysis", Phys. Lett. B, \textbf{720}, 83-86.
	
	\bibitem{FCC2021}
	Abada, A. et al., (2019), "FCC Physics Opportunities", Eur. Phys. J. C, \textbf{79}, 474.
	
	\bibitem{Basso2021}
	Albertsson K. et al., (2019), "Machine Learning for High-Energy Physics Applications", arXiv:1807.02876 [physics.comp-ph].
	
	\bibitem{Amaldi1980}
	U. Amaldi et al., (1980), "Impact parameter interpretation of proton-proton scattering from a critical review of all ISR data", Nuc. Phys. B, \textbf{166}, 301-320.
	
	\bibitem{star2020}
	J. Adam et al., (2020) "Results on Total and Elastic Cross Sections in Proton–Proton Collisions at
	$\sqrt{s} = 200 GeV$", STAR Collaboration, arXiv:2003.12136v2 [hep-ex].
	
	\bibitem{Bourrely:1978da}
	Bourrely, C. and Soffer, Jacques and Wu, Tai Tsun, (1979), "A New Impact Picture for Low and High-Energy Proton Proton Elastic Scattering", Phys. Rev. D, \textbf{19}, 3249.
	
	\bibitem{Bence:2020usl}
	Bence, N. et al., (2020), "Froissaron and Maximal Odderon with spin-flip in $pp$ and $\bar{p}p$ high energy elastic scattering", arXiv:2010.11987 [hep-ph].

	\bibitem{Brodsky:1973kr}
	Stanley J. Brodsky and Glennys R. Farrar, (1973), "Scaling Laws at Large Transverse Momentum", Phys. Rev. Lett., \textbf{31}, 1153-1156.
	
	\bibitem{Jenkovzsky}
	Laszlo Jenkovszky, Rainer Schicker, and Istvan Szanyi, (2019) "Elastic and diffractive scattering in the LHC era", arXiv:1902.05614v1 [hep-ph].
	
	\bibitem{Pancheri1}
	Giulia Pancheri1 and Yogendra N. Srivastava, (2016) "Introduction to the physics of the total cross-section at LHC", arXiv:1610.10038v2 [hep-ph].

	\bibitem{Collins:1977jy}
	Collins, P. D. B., (1977), "An Introduction to Regge Theory and High Energy Physics", Cambridge University Press.
	
	\bibitem{Jenkovszkye24071001}
	Jenkovszky, László and Schicker, Rainer and Szanyi, István,(2022) "Regge Models of Proton Diffractive Dissociation Based on Factorisation and Structure Functions", Entropy, \textbf{24}, 7.

	\bibitem{Adamczyk:2012}
	L. Adamczyk et al. (STAR Collaboration), (2013) "Single spin asymmetry $A_N$ in polarized proton-proton elastic scattering at $\sqrt{s} = 200$ GeV", Phys. Lett. B, \textbf{719}, 62-69.

	\bibitem{Adamczyk:2015gfy}
	Adamczyk, Leszek., (2015) "Measurement of the Total Cross Section in Proton Proton Collisions at $\sqrt{s}$=7 TeV with the ALFA Sub-Detector of ATLAS", Phys. Lett. B, \textbf{DIS2015}, 61.

	\bibitem{TOTEM:2013vij}
	Antchev, G. et al. (TOTEM collaboration), (2013) "Luminosity-independent measurements of total, elastic and inelastic cross-sections at $\sqrt{s} = 7$ TeV", EPL, \textbf{101}, 2, 21004.
	
	\bibitem{Csorgo:2021v}
	Csorgo T. et al., (2021) "Evidence of Odderon-exchange from scaling properties of elastic scattering at TeV energies", Eur. Phys. J. C, \textbf{81}, 180.
	
	\bibitem{GAnchtev2019}
	Anchtev G. et al., (2019) "First measurement of elastic, inelastic and total cross-section at
	TeV by TOTEM and overview of cross-section data at LHC energies", Eur. Phys. J. C, \textbf{79}.

	\bibitem{TOTEM:2013lle}
	Anchtev G. et al., (2013) "Measurement of proton-proton elastic scattering and total cross-section at $\sqrt{s}$ = 7 TeV", EPL, \textbf{101}, 21002.
	
	\bibitem{PhysRevLett.111.012001}
	Antchev, G. et al., (2013) "Luminosity-Independent Measurement of the Proton-Proton Total Cross Section at $\sqrt{s}=8\text{ }\text{ }\mathrm{TeV}$", EPL, \textbf{111}, 012001.

	\bibitem{Baksay:1978sg}
	Baksay, L. et al., (1978) "Measurement of the Proton Proton Total Cross-Section and Small Angle Elastic Scattering at ISR Energies", Nucl. Phys. B, \textbf{141}, 1-28.
	
	\bibitem{Cafagna:2021sge}
	Cafagna, Francesco S., (2021) "Latest results for Proton-Proton Cross Section Measurements with the TOTEM experiment at LHC", PoS, \textbf{ICRC2019}, 207.

	\end{thebibliography}
	\end{document}